\documentclass{article}
\usepackage[final]{graphics}
\usepackage{amsfonts,amsbsy}

\def\empile#1\over#2{\mathrel{\mathop{\kern 0pt#1}\limits_{#2}}}

\font\tenmsa=msam10
\font\sevenmsa=msam7
\font\fivemsa=msam5
\font\tenmsb=msbm10
\font\sevenmsb=msbm7
\font\fivemsb=msbm5
\newfam\msafam
\newfam\msbfam
\textfont\msafam=\tenmsa  \scriptfont\msafam=\sevenmsa
\scriptscriptfont\msafam=\fivemsa
\textfont\msbfam=\tenmsb  \scriptfont\msbfam=\sevenmsb
\scriptscriptfont\msbfam=\fivemsb

\catcode`\@=11

\newcount\@tempcntc
\def\@citex[#1]#2{\if@filesw\immediate\write\@auxout{\string\citation{#2}}\fi
  \@tempcnta\z@\@tempcntb\m@ne\def\@citea{}\@cite{%
        \@for\@citeb:=#2\do%
    {\@ifundefined{b@\@citeb}%
        {\@citeo\@tempcntb\m@ne\@citea%
                \def\@citea{,\penalty\@m\ }{\bf ?}\@warning%
                {Citation `\@citeb' on page \thepage \space undefined}}%
        {\setbox\z@\hbox{\global\@tempcntc0\csname b@\@citeb\endcsname\relax}
     \ifnum\@tempcntc=\z@ \@citeo\@tempcntb\m@ne%
       \@citea\def\@citea{,\penalty\@m}%
       \hbox{\csname b@\@citeb\endcsname}%
     \else%
      \advance\@tempcntb\@ne%
      \ifnum\@tempcntb=\@tempcntc%
      \else\advance\@tempcntb\m@ne\@citeo%
      \@tempcnta\@tempcntc\@tempcntb\@tempcntc\fi\fi}}\@citeo}{#1}}%

\def\@citeo{\ifnum\@tempcnta>\@tempcntb\else\@citea
  \def\@citea{,\penalty\@m}%
  \ifnum\@tempcnta=\@tempcntb\the\@tempcnta\else
   {\advance\@tempcnta\@ne\ifnum\@tempcnta=\@tempcntb \else
\def\@citea{--}\fi
    \advance\@tempcnta\m@ne\the\@tempcnta\@citea\the\@tempcntb}\fi\fi}

\catcode`\@=12

\global\mathchardef\lesssim "142E

\newcommand{\slL}{\raise.15ex\hbox{$/$}\kern-.53em\hbox{$L$}}
\newcommand{\slP}{\raise.15ex\hbox{$/$}\kern-.53em\hbox{$P$}}
\newcommand{\slp}{\raise.1ex\hbox{$/$}\kern-.63em\hbox{$p$}}
\newcommand{\slq}{\raise.1ex\hbox{$/$}\kern-.63em\hbox{$q$}}
\newcommand{\slv}{\raise.1ex\hbox{$/$}\kern-.63em\hbox{$v$}}
\newcommand{\slR}{\raise.15ex\hbox{$/$}\kern-.53em\hbox{$R$}}
\newcommand{\slQ}{\raise.15ex\hbox{$/$}\kern-.53em\hbox{$Q$}}
\newcommand{\slK}{\raise.15ex\hbox{$/$}\kern-.53em\hbox{$K$}}
\newcommand{\slk}{\raise.15ex\hbox{$/$}\kern-.53em\hbox{$k$}}
\newcommand{\slSigma}{\raise.15ex\hbox{$/$}\kern-.53em\hbox{$\Sigma$}}
\newcommand{\slcalP}{\raise.15ex\hbox{$/$}\kern-.63em\hbox{$\cal P$}}
\newcommand{\slA}{\raise.15ex\hbox{$/$}\kern-.73em\hbox{$A$}}
\newcommand{\slbfA}{\raise.15ex\hbox{$/$}\kern-.73em\hbox{${\imb A}$}}
\newcommand{\slpartial}{\raise.15ex\hbox{$/$}\kern-.53em\hbox{$\partial$}}

\newcommand{\be}{\begin{equation}}
\newcommand{\ee}{\end{equation}}
\newcommand{\bea}{\begin{eqnarray}}
\newcommand{\ena}{\end{eqnarray}}

\def\build#1\over#2{\mathrel{\mathop{\kern 0pt#1}\limits_{#2}}}

\font\tenimbf=cmmib10 at 10pt
\font\sevenimbf=cmmib10 at 7pt
\font\fiveimbf=cmmib10 at 5pt
\newfam\imbf
\textfont\imbf=\tenimbf
\scriptfont\imbf=\sevenimbf
\scriptscriptfont\imbf=\fiveimbf
\def\imb{\fam\imbf\tenimbf}

\begin{document}
\title{\bf{Enhanced thermal production of hard dileptons by $3\to 2$
    processes}} \author{P.~Aurenche$^{(1)}$, F.~Gelis$^{(2)}$,
  H.~Zaraket$^{(3)}$} \maketitle
\begin{center}
\begin{enumerate}
\item Laboratoire d'Annecy-le-Vieux de Physique Th\'eorique,\\
Chemin de Bellevue, B.P. 110,\\
74941 Annecy-le-Vieux Cedex, France
\item Laboratoire de Physique Th\'eorique,\\
B\^at. 210, Universit\'e Paris XI,\\
91405 Orsay Cedex, France
\item Physics Department and Winnipeg Institute
for Theoretical Physics,\\
University of Winnipeg,\\
Winnipeg, Manitoba R3B 2E9, Canada
\end{enumerate}
\end{center}

\begin{abstract}
  In the framework of the Hard Thermal Loop effective theory, we
  calculate the two-loop contributions to hard lepton pair production
  in a quark-gluon plasma.  We show that the result is free of any
  infrared and collinear singularity. We also recover the known fact
  that perturbation theory leads to integrable singularities at the
  location of the threshold for $q\bar{q}\to\gamma^*$. It appears that
  the process calculated here significantly enhances the rate of low
  mass hard dileptons.
\end{abstract}
\vskip 4mm
\centerline{\hfill LAPTH-908/02, LPT-ORSAY-02/28}

\section{Introduction}
\begin{figure}
\centerline{\resizebox*{!}{3cm}{\includegraphics{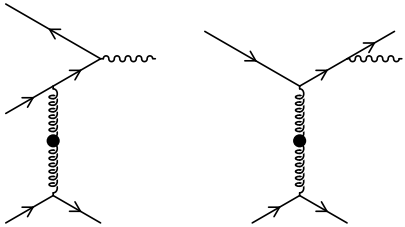}}}
\caption{\label{fig:processes} The bremsstrahlung and off-shell
annihilation processes that are expected to dominate the production of
low mass hard dileptons.}
\end{figure}
Some time ago a new mechanism for the production of hard photons in a
quark-gluon plasma was proposed \cite{AurenGKZ1}: the photon is produced
by the annihilation of a quark-antiquark pair ``after" one of the
annihilating quark or antiquark has scattered in the medium (diagram on
the right of figure \ref{fig:processes}). This process is contained in
the two-loop diagrams \cite{AurenGKP1,AurenGKP2} of the Hard Thermal
Loop (HTL - \cite{Pisar6,BraatP1,BraatP2,FrenkT1,FrenkT2}) resummed
theory. In the HTL approach quarks acquire a thermal mass in the plasma
and, at lowest order in the strong coupling, the annihilation of a
quark-antiquark pair into a real photon is kinematically forbidden.
However when the annihilation is associated to scattering in the plasma
one of the quarks becomes off-shell and the annihilation is
possible. The interesting feature of the mechanism is that, apart from
the exponential suppression factor common to all thermal processes, the
photon production rate grows with the energy of the photon, leaving it
as the only relevant mechanism when the photon energy is much larger
than the temperature. In particular, it dominates over all other
processes contained in the one-loop approximation
\cite{BraatPY1,BaierNNR1,KapusLS1,BaierPS1,AurenBP1,Niega6} of the HTL
approach ($qg\to q\gamma$ and $q \bar q\to g\gamma$) which only grow
logarithmically with the photon energy. Recent studies, in the framework
of the hydrodynamical evolution of the plasma
\cite{Sriva1,SrivaS1,HuoviRR1,AlamSRHS1,AlamSHNS1}, have shown that
thermal production of a photon, including the ``off-shell annihilation"
process, is already significant at SPS energies and could dominate at
RHIC and LHC energies. There is then a window, of photon energies of a
few GeV, which could provide a signal for the formation of the plasma in
heavy ion collisions. However, in this energy range, there is a very
large background of pions which decay into photons.  Whether the flux of
thermal photons can be identified over the large background is under
study. In view of this problem it may be useful to consider an
alternative signal which follows the same dynamics as real photon
production but which suffers from a different experimental
background. The production of small mass ($\sqrt{Q^2}$) dilepton pairs
at relatively hard energy ($q_0$) is such a process and this is what is
considered below in the limit $\sqrt {Q^2} \ll q_0$.  Considering lepton
pair production leads to an interesting situation where two hard scales,
$T$ the temperature and $q_0$, and three small scales, $\sqrt {Q^2}$,
$M_\infty \sim g T$ the thermal mass of fermions and $m_{\rm g} \sim gT$ the
thermal mass of the gluon enter the game. It turns out that several of
these scales combine in a single expression $M_{\rm{eff}}$ which acts as
a cut-off to regularize potential collinear singularities associated
with the fermion propagators.  However, in the off-shell annihilation
process, $M^2_{\rm{eff}}$ can vanish and even become negative. Its role
as a cut-off is then blurred! This occurs when the condition
$$
Q^2 > 4 \ M^2_{\infty}$$ is satisfied. In this case, lepton pairs can be
produced at ${\cal O}(0)$ in the strong interactions by the annihilation
of a $q \bar q$ pair, a new feature compared to the case of real photon
production. When going to higher orders in the strong interactions it
means that two types of diagrams will contribute, which are usually
referred to as real and virtual diagrams. This point, which involves the
compensation of divergences between the two types of diagrams
\cite{Kinos1,LeeN1,BaierPS2,AltheAB1,GabelGP1,MajumG1}, will be
discussed in some details below as it has no equivalent in the case of
real photon production. Even after all divergences are cancelled, the
dilepton invariant mass spectrum still exhibits an integrable
singularity at $Q^2 = 4 \ M^2_{\infty}$. This is in agreement with a
theorem of perturbation theory \cite{CatanW1} which states that if an
observable has its phase space restricted by a $\theta$ function at a
given order of the perturbative expansion, it may exhibit $\delta$
function singularity at the next order.

The collinear enhancement mechanism that makes the processes of figure
\ref{fig:processes} important has also been shown to play a role in
multi-loop diagrams belonging to the class of ladder corrections and
self-energy-corrections
\cite{LebedS1,LebedS2,AurenGZ2,Gelis11,Gelis12,ArnolMY1,ArnolMY2}, and a
resummation of this family of diagrams has been carried out in
\cite{ArnolMY1,ArnolMY2}. The effect of this resummation, also known as
Landau-Pomeranchuk-Migdal (LPM) effect \cite{LandaP1,LandaP2,Migda1},
leads to a small reduction (by about 25\% for photons in the range
interesting for phenomenology) of the photon rate. Assuming that the
magnitude of this effect remains comparable for reasonable photon
invariant masses (the photon mass helps to regularize the collinear
singularities), we limit ourselves to a 2-loop calculation. These
contributions add up to the processes $q\bar{q}\to g\gamma^*$ and
$qg\to q\gamma^*$ already calculated in
\cite{AltheR1,ThomaT2}.

In the following section, we set up the formalism, then we discuss the
matrix element showing, before carrying out the phase space
integration, that cancellation of some infra-red divergences can be
expected between vertex and self energy diagrams.  In Sec. 4 the phase
space integration is carried out while in Sec. 5 the behavior of the
spectrum near the singular point $Q^2 = 4 \ M^2_{\infty}$ is studied
analytically and found to be in complete agreement with the numerical
studies. Finally, Sec. 6 is devoted to some simple phenomenological
studies.  An appendix contains details of the phase space integration.

In our study on dilepton pair production we rederive, as the limit when
the dilepton mass vanishes, the results already obtained on real photon
production. The method we follow is however simpler than previously
used. It still uses the retarded/advanced (R/A) formalism
\cite{AurenB1,AurenBP1,EijckW1,EijckKW1,Gelis3} but it relies on
carrying out some integrations in the complex energy plane for which the
R/A formalism is well suited. Furthermore, when discussing the
compensation of divergences and calculating the remaining finite pieces,
there is no need to go to $n-$dimensions or introduce regulators since
the compensations can be seen within the integrand.

Some of the results of this paper rely on a new sum-rule presented
elsewhere \cite{AurenGZ4}, which allows to analytically carry out
otherwise difficult numerical integrations.

\section{Model}
\subsection{Formula for the dilepton rate}
In a quark-gluon plasma in local thermal equilibrium, the number of
lepton pairs produced from virtual photons per unit time and per unit
volume is conveniently expressed in terms of the imaginary part of the
retarded photon polarization tensor, calculated in presence of the
plasma \cite{Weldo3,GaleK1}:
\begin{equation}
{{dN^{l^+l^-}}\over{dtd^3{\imb x}}}=-{{d\omega d^3{\imb
q}}\over{12\pi^4}} {\alpha\over{Q^2}} n_{_{B}}(\omega)\, {\rm
Im}\,\Pi_{_{R}}{}^\mu_\mu(\omega,{\imb q})\; .
\label{eq:def-rate}
\end{equation}
In this formula, the sum over the phase-space of the lepton pair in
which the virtual photon decays has been performed, at a fixed total
momentum $Q\equiv(\omega,{\imb q})$ of the pair. The trace over the
Lorentz indices $\cdots{}^\mu_\mu$ indicates that both the
longitudinal and transverse modes of the massive photon are included.
Note that this formula is valid at lowest order in the electromagnetic
coupling constant, and at all orders in the strong coupling constant.
In more physical terms, it neglects the possible reinteractions of the
photon or leptons on their way out of the plasma, and it is valid only if
the size of the system is small compared to the mean free path of a
photon or lepton.

Another limitation of this formula is that it holds for
a plasma in thermal and chemical equilibrium. However, it can also be
applied to a situation where there is only a local equilibrium, provided
the typical size of the cells over which the system can be considered to
be equilibrated is large in front of the photon formation
time\footnote{This is the implicit assumption in the hydrodynamical
approach used in order to describe photon emission during the collision
of two heavy ions.} \cite{Gelis11,GelisSS1}. If this condition is not
satisfied, one has to go back to basics, and use a non-equilibrium
real-time formulation in order to compute the rate
\cite{WangBN1,BoyanVW1,Dadic1,Dadic2,Dadic3} \footnote{Although the
assumption of an infinitely fast switching of the coupling constant that
has been made in those calculations may lead to unphysical effects like
a power-like tail in the emission spectrum \cite{Moore1}.}.

\subsection{Hard thermal loops}
In order to perform this calculation using thermal field theory, one
must resum the bare propagators and vertices in order to account for the
fact that medium effects modify the interactions and properties of the
excitations of the plasma. In particular, medium induced masses are of
utmost importance for quarks in processes where the photon is
predominantly produced in a collinear configuration. In thermal field
theory, this is achieved through the HTL resummation, which resums an
infinite set of one-loop thermal corrections to propagators and
vertices.

Since the quarks are always hard in the processes we consider throughout
this paper, we keep only the asymptotic thermal mass from the HTL
fermionic self-energy \cite{FlechR1}, which leads to the following
retarded and advanced propagators
\begin{equation}
S_{_{R,A}}(P)\equiv \overline{\slP}s_{_{R,A}}(P)\; ,
\end{equation}
with
\begin{eqnarray}
&&\overline{P}\equiv(p_0,\sqrt{{\imb p}^2+M_\infty^2}\hat{\imb p})\nonumber\\
&&
s_{_{R,A}}(P)={i\over{P^2-M_\infty^2\pm i p_0\epsilon}}\; ,
\label{eq:eff-prop}
\end{eqnarray}
where $M_\infty^2\equiv g^2 C_{_{F}} T^2/4$ is the square of the thermal
mass of a hard quark ($C_{_{F}}\equiv (N_c^2-1)/2N_c$ is the Casimir in
the fundamental representation of the gauge group
$SU(N_c)$). Ordinarily, one does not need effective $\gamma\bar{q}q$
vertices if the quarks are hard, since the HTL corrections are
suppressed. But this is correct only if the terms we are calculating are at
leading order. Photon production however involves an important
cancellation between the components $\Pi_{00}$ and $\Pi_{zz}$ of the
photon polarization tensor, and the remaining terms in $\Pi_{\mu}{}^\mu$
are non leading, and are affected by the HTL corrections to the
vertex\footnote{This vertex correction has never been considered in the
existing calculations of photon production. As we shall see later, it
slightly modifies the value of $\Pi_{\mu}{}^\mu$. However, one can check
that it modifies only the $\Pi_{zz}$ component, and therefore forgetting
it had no impact on the result of \cite{ArnolMY1,ArnolMY2} who
calculated only the transverse components (since $\Pi_{00}=\Pi_{zz}$ for
real photons). For virtual photons, one has to include the temporal and
longitudinal components as well, and one cannot avoid using this vertex
correction.}. In the following, we do not need the explicit form of the
vertex, but only need to know how it is related to the self-energy
correction by means of Ward identities. In other words, it is important
to make sure that we do for the effective vertex an approximation which
is consistent with the approximation of Eq.~(\ref{eq:eff-prop}) made for
the quark propagator. In particular, we want to preserve the relations
satisfied by the full HTL corrections (for a vertex in which a quark
enters with momentum $R$ and goes out with momentum $P$):
\begin{eqnarray}
&&P_\mu\delta\Gamma^\mu=\slSigma(R)=\slR-\overline{\slR}\; ,\nonumber\\
&&R_\mu\delta\Gamma^\mu=\slSigma(P)=\slP-\overline{\slP}\; ,
\label{eq:ward}
\end{eqnarray}
where $\delta\Gamma$ is the HTL correction to the vertex. In the
following, it will be useful to write
\begin{equation}
\delta\Gamma^\mu\equiv\Gamma^{\mu\nu}\gamma_\nu\; .
\end{equation}

In addition, we need the effective retarded and advanced propagators
of a gluon, which can be decomposed into its transverse and
longitudinal components:
\begin{equation}
-D^{\mu\nu}_{_{R,A}}(L)\equiv
 P_{_{T}}^{\mu\nu}(L) \Delta_{_{R,A}}^{^{T}}(L)
+P_{_{L}}^{\mu\nu}(L) \Delta_{_{R,A}}^{^{L}}(L)\; ,
\end{equation}
where $P_{_{T,L}}^{\mu\nu}(L)$ are respectively the transverse and
longitudinal projectors\footnote{They satisfy the following useful
  identities:
\begin{eqnarray}
&&P_{_{T}} P_{_{T}}=P_{_{T}}\; ,\qquad
P_{_{L}} P_{_{L}}=P_{_{L}}\; ,\qquad
P_{_{T}}P_{_{L}}=P_{_{L}}P_{_{T}}=0\; ,\nonumber\\
&&L_\mu P_{_{T,L}}^{\mu\nu}(L)=0\; .
\end{eqnarray}} for a gauge boson of momentum $L$ \cite{LandsW1}:
\begin{eqnarray}
&&P^{\mu\nu}_{_{T}}(L)=\gamma^{\mu\nu}-{{\kappa^\mu\kappa^\nu}
\over{\kappa^2}}\nonumber\\
&&P^{\mu\nu}_{_{L}}(L)=g^{\mu\nu}-{{L^\mu L^\nu}\over{L^2}}-P_{_{T}}^{\mu\nu}(L)\; ,
\end{eqnarray}
where $\gamma^{\mu\nu}\equiv g^{\mu\nu}-U^\mu U^\nu$, $\kappa^\mu \equiv
\gamma^{\mu\nu}L_\nu$, and where $U$ is the mean velocity of the plasma
in the frame under consideration.  The propagators of the transverse and
longitudinal modes are then given by
\begin{equation}
\Delta_{_{R,A}}^{^{T,L}}(L)\equiv
{i\over{L^2-\Pi_{_{R,A}}^{^{T,L}}(L)}}\; ,
\end{equation}
with the following HTL self-energies
\begin{eqnarray}
&\Pi^{^{T}}(L)=3 m_{\rm g}^2&\left[ {{x^2}\over
2}+{{x(1-x^2)}\over{4}}\ln\left({{x+1}\over{x-1}}\right)
\right]\nonumber\\ 
&\Pi^{^{L}}(L)=3 m_{\rm g}^2&\left[
(1-x^2)-{{x(1-x^2)}\over{2}}\ln\left({{x+1}\over{x-1}}\right)
\right]\; ,\nonumber\\ &&
\label{gluonself}
\end{eqnarray}
where we denote $x\equiv l_0/l$ and where $m_{\rm g}^2\equiv g^2 T^2 [N_c+
N_{_{F}}/2]/9$ is the gluon thermal mass in an $SU(N_c)$ gauge theory
with $N_{_{F}}$ flavors. Note that an implicit continuation $l_0\to
l_0\pm i\epsilon$ is understood in this self-energy in order to obtain
its retarded and advanced components.

\section{Matrix element}
\subsection{Diagrams}
\begin{figure}
\centerline{\resizebox*{!}{4cm}{\includegraphics{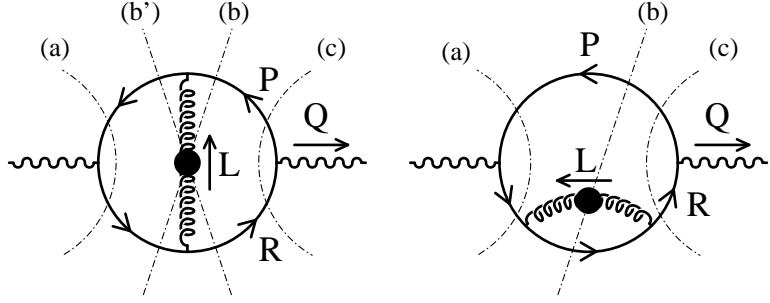}}}
\caption{\label{fig:diagrams} The two-loop diagrams contributing to
  the production of hard dileptons. A third diagram with a self-energy
  insertion on the upper quark line is not shown here.}
\end{figure}

Let us now move to the two-loop diagrams calculated in this
paper. They are the same as those already considered in
\cite{AurenGKP1,AurenGKP2,AurenGKZ1} for real photons and in
\cite{AurenGKZ2} for soft dileptons produced at rest in the plasma
frame. One of them is a vertex correction to the one-loop photon
self-energy, and the other is a self-energy insertion on one of the
quarks lines (see figure \ref{fig:diagrams}).

There is however one major difference compared to the case of real
photons: as long as the invariant mass of the lepton pair is $Q^2<
4M_\infty^2$, the cuts denoted by $(a)$ and $(c)$ in figure
\ref{fig:diagrams} do not contribute. Indeed, they correspond to the
interference between the tree-level process $q\bar{q}\to \gamma^*$ and a
virtual correction to this process, but this direct production mechanism
has a threshold at $Q^2=4M_\infty^2$. Above the threshold, one must
include these cuts, and in fact the various cuts partly compensate each
other through the KLN mechanism.

\subsection{Vertex contribution}
It is convenient to use the retarded-advanced formalism
\cite{AurenB1,AurenBP1,EijckW1,EijckKW1,Gelis3} in order to compute the
retarded photon polarization tensor needed in
Eq.~(\ref{eq:def-rate}). Throughout this paper, we assume that the gluon
of momentum $L$ (see diagrams and notations in figure
\ref{fig:diagrams}) is soft, so that the corresponding Bose-Einstein
factor $n_{_{B}}(l_0)$ is large, in particular large compared to
fermionic statistical factors. When performing the integrals, one can
check the validity of this hypothesis. Therefore, we keep only those terms
containing this enhanced Bose-Einstein factor, and we approximate
$n_{_{B}}(l_0)\approx T/l_0$.

Using this approximation, the contribution of the first diagram (which
we designate by ``vertex correction'' in the following) of figure
\ref{fig:diagrams} to the photon polarization tensor is given by:
\begin{eqnarray}
&&{\rm Im}\,\Pi_{_{R}}{}_\mu^\mu(Q)\Big|_{\rm vertex}\approx
{1\over 2} e^2 g^2 N_c C_{_{F}}
\int{{d^4P}\over{(2\pi)^4}}[n_{_{F}}(r_0)-n_{_{F}}(p_0)]\nonumber\\
&&\qquad\qquad\qquad\qquad\times\int {{d^4L}\over{(2\pi)^4}}{T\over{l_0}}
\sum_{\alpha=T,L}
\rho_\alpha(L) P_\alpha^{\rho\sigma}(L) {\rm Trace}^{\rm
vert}_{\rho\sigma}\nonumber\\
&&\qquad\qquad\qquad\qquad\times
[s_{_{R}}(P)s_{_{R}}(P+L)-s_{_{A}}(P)s_{_{A}}(P+L)]\nonumber\\
&&\qquad\qquad\qquad\qquad\times
[s_{_{R}}(R)s_{_{R}}(R+L)-s_{_{A}}(R)s_{_{A}}(R+L)]\; ,
\label{eq:vertex1}
\end{eqnarray}
where $\rho_\alpha(L)$ is the spectral function for the mode $\alpha$ of
the effective gluon propagator:
\begin{equation}
\rho_\alpha(L)\equiv \Delta_{_{R}}^\alpha(L)-\Delta_{_{A}}^\alpha(L)\; ,
\end{equation}
and where ${\rm Trace}^{\rm vert}_{\rho\sigma}$ denotes the trace of
the Dirac matrices for this diagram:
\begin{equation}
{\rm Trace}^{\rm vert}_{\rho\sigma}\equiv{\rm Tr}\,[\overline{\slP}(\gamma^\mu+\delta\Gamma^\mu)\overline{\slR}
(\gamma_\rho+\delta\Gamma_\rho)(\overline{\slR+\slL})(\gamma_\mu+\delta\Gamma_\mu)
(\overline{\slP+\slL})(\gamma_\sigma+\delta\Gamma_\sigma)]\, .
\end{equation}
This trace contains four hard momenta, and is a priori of order $T^4$ or
larger. It is however important to keep also subleading terms of order
$T^2 M_\infty^2$ since the cancellation between $\Pi_{00}$ and
$\Pi_{zz}$ kills all the leading terms when the vertex and self-energy
diagrams are combined. However, it is safe to drop all the smaller
corrections, i.e. terms of order $M_\infty^4$ or smaller. Expanding this trace
up to terms with two powers of a soft quantity, we
find\footnote{Anticipating the fact that $L^\rho$ contracted with the
projectors $P_{_{T,L}}^{\rho\sigma}(L)$ gives a vanishing result, we
have dropped the terms proportional to this quantity.}:
\begin{eqnarray}
{\rm Trace}^{\rm vert}_{\rho\sigma}&\approx&
-4\Big[2(\overline{R}^2+(\overline{R+L})^2)P_\rho Q_\sigma
-2(\overline{P}^2+(\overline{P+L})^2)R_\rho Q_\sigma\nonumber\\
&&\qquad+2 L^2 (R_\rho R_\sigma+P_\rho P_\sigma)
+4(\overline{P}^2+\overline{R}^2)P_\rho R_\sigma\nonumber\\
&&\qquad-8(\overline{P}^\alpha+\Gamma^{\alpha\beta}P_\beta)
(\overline{R}_\alpha+\Gamma_{\alpha}{}^\gamma R_\gamma)P_\rho R_\sigma
\nonumber\\
&&\qquad-8(Q^\alpha+\Gamma^{\alpha\beta}Q_\beta)
(Q_\alpha+\Gamma_{\alpha}{}^\gamma Q_\gamma)P_\rho R_\sigma
\Big]\; .
\end{eqnarray}
In this expression, one can even drop the terms with two powers of the
vertex correction $\Gamma^{\mu\nu}$, since they would be smaller than
the terms we are interested in here.  Using the pseudo Ward identities
of Eqs.~(\ref{eq:ward}), it is possible to rewrite the three terms in
$P_\rho R_\sigma$ as
\begin{equation}
16\left[Q^2+2(P\cdot R-\overline{P}\cdot\overline{R})\right]P_\rho
R_\sigma\; ,
\label{eq:trace1}
\end{equation}
up to terms that contain more powers of soft factors.

\subsection{Self-energy contribution}
Similarly, the contribution of the diagram with a self-energy insertion
on the quark of momentum $R$ is given by
\begin{eqnarray}
&&{\rm Im}\,\Pi_{_{R}}{}_\mu^\mu(Q)\Big|_{\rm self}\approx
{1\over 2} e^2 g^2 N_c C_{_{F}}
\int{{d^4P}\over{(2\pi)^4}}[n_{_{F}}(r_0)-n_{_{F}}(p_0)]\nonumber\\
&&\qquad\qquad\qquad\qquad\times\int {{d^4L}\over{(2\pi)^4}}{T\over{l_0}}
\sum_{\alpha=T,L}
\rho_\alpha(L) P_\alpha^{\rho\sigma}(L) {\rm Trace}^{\rm
self}_{\rho\sigma}\nonumber\\
&&\qquad\qquad\qquad\qquad\times
[s_{_{R}}(P)-s_{_{A}}(P)]\nonumber\\
&&\qquad\qquad\qquad\qquad\times
[[s_{_{R}}(R)]^2s_{_{R}}(R+L)-[s_{_{A}}(R)]^2s_{_{A}}(R+L)]\; ,
\label{eq:self1}
\end{eqnarray}
where ${\rm Trace}^{\rm self}_{\rho\sigma}$ is the corresponding Dirac
trace:
\begin{eqnarray}
{\rm Trace}^{\rm self}_{\rho\sigma}\equiv {\rm
Tr}\,[\overline{\slP}(\gamma^\mu+\delta\Gamma^\mu)\overline{\slR}(\gamma_\rho+\delta\Gamma_\rho)(\overline{\slR+\slL})(\gamma_\sigma+\delta\Gamma_\sigma)\overline{\slR}(\gamma_\mu+\delta\Gamma_\mu)]\; .
\end{eqnarray}
Calculating this trace down to the order $T^2M_\infty^2$, we obtain
\begin{eqnarray}
{\rm Trace}^{\rm self}_{\rho\sigma}&\approx& -4\Big[
4\overline{R}^2 R_\rho Q_\sigma 
+4(\overline{P}^2+\overline{R}^2)R_\rho R_\sigma\nonumber\\
&&\qquad-8(\overline{P}^\alpha+\Gamma^{\alpha\beta}P_\beta)
(\overline{R}_\alpha+\Gamma_{\alpha}{}^\gamma R_\gamma)R_\rho R_\sigma
\nonumber\\
&&\qquad-8(Q^\alpha+\Gamma^{\alpha\beta}Q_\beta)
(Q_\alpha+\Gamma_{\alpha}{}^\gamma Q_\gamma)R_\rho R_\sigma
\Big]\; .
\label{eq:self2}
\end{eqnarray}
We can recognize the same set of terms as the one evaluated in
Eq.~(\ref{eq:trace1}), with $P_\rho R_\sigma$ replaced by $R_\rho
R_\sigma$.  Of course, there is also another such diagram, not
represented explicitly in figure \ref{fig:diagrams}, with the
self-energy insertion on the quark of momentum $P$. In fact, the
contribution of this diagram is equal to that of the first self-energy
insertion. Alternatively, the corresponding matrix element can be
obtained from the previous one by performing the following change of
variables:
\begin{eqnarray}
&&Q\to Q\; ,\nonumber\\
&&R \to -P\; ,\nonumber\\
&&L\to -L\; .
\end{eqnarray}
This second method, compared to the one where we just multiply
Eq.~(\ref{eq:self2}) by a factor 2, has the advantage of making the
matrix element symmetric under the exchange of $P$ and $R$ and is
preferred in order to see some cancellations. 

\subsection{Partial cancellation between vertex and self}
Without performing any explicit calculation, one can first prove that
the terms in $R_\rho Q_\sigma$ and $P_\rho Q_\sigma$ all cancel when
one combines the contribution of the vertex correction diagram, and of
the two self-energy insertion diagrams.  These terms can
be rewritten as\footnote{The complex poles of the statistical factors
  do not contribute since the difference between retarded and advanced
  propagators in Eqs~(\ref{eq:vertex1}) and (\ref{eq:self1}) vanishes
  when evaluated at such poles.}
\begin{eqnarray}
&&-2 e^2 g^2 N_c C_{_{F}}
\int{{d^4P}\over{(2\pi)^4}}[n_{_{F}}(r_0)-n_{_{F}}(p_0)]\nonumber\\
&&\qquad\qquad\times\int {{d^4L}\over{(2\pi)^4}}{T\over{l_0}}
\sum_{\alpha=T,L}
\rho_\alpha(L) P_\alpha^{\rho\sigma}(L)\nonumber\\
&&\qquad\qquad\times\Big[
2(\overline{R}^2+(\overline{R+L})^2)P_\rho Q_\sigma\;
s_{_{R}}(P)s_{_{R}}(P+L)
s_{_{A}}(R)s_{_{A}}(R+L)\nonumber\\
&&\qquad\qquad\quad-2(\overline{P}^2+(\overline{P+L})^2)R_\rho Q_\sigma\;
s_{_{R}}(P)s_{_{R}}(P+L)
s_{_{A}}(R)s_{_{A}}(R+L)\nonumber\\
&&\qquad\qquad\quad+4\overline{R}^2 R_\rho Q_\sigma\;
 s_{_{R}}(P)
[s_{_{A}}(R)]^2s_{_{A}}(R+L)\nonumber\\
&&\qquad\qquad\quad-4\overline{P}^2 P_\rho Q_\sigma\;
 s_{_{A}}(R)
[s_{_{R}}(P)]^2s_{_{R}}(P+L)
\Big]+{\rm c.c.}\; .
\end{eqnarray}
Using $\overline{P}^2s_{_{R,A}}(P)=i$, one can check that the square
bracket in this equation is proportional to:
\begin{eqnarray}
&&\Big[
4 P_\rho Q_\sigma\; s_{_{R}}(P)s_{_{R}}(P+L)
\Big(s_{_{A}}(R+L)-s_{_{A}}(R)\Big)
\nonumber\\
&&-4 R_\rho Q_\sigma\; s_{_{A}}(R)s_{_{A}}(R+L)
\Big(s_{_{R}}(P+L)-s_{_{R}}(P) \Big)
\Big]\; .
\end{eqnarray}
Assuming that the gluon momentum $L$ is small compared to the quark
momenta $P$ and $R$, and performing the change of variables
\begin{eqnarray}
&&P+L\to -R\; ,\nonumber\\
&&L\to L\; ,
\label{eq:trans1}
\end{eqnarray}
on some of the terms (dropping terms in $L_\rho$ that may appear), one
can observe that the two lines cancel each other. This cancellation is
in fact a particular case of a cancellation between ladder diagrams and
self-energy insertions discussed in
\cite{LebedS1,LebedS2,CarriKP1,CarriK1,Bodek1,AurenGZ2,ArnolMY1,ArnolMY2}. For
the remaining terms, we have:
\begin{eqnarray}
&&{\rm Im}\,\Pi_{_{R}}{}_\mu^\mu(Q)\approx 
-2 e^2 g^2 N_c C_{_{F}}
\int{{d^4P}\over{(2\pi)^4}}[n_{_{F}}(r_0)-n_{_{F}}(p_0)]\nonumber\\
&&\qquad\times\int {{d^4L}\over{(2\pi)^4}}{T\over{l_0}}
\sum_{\alpha=T,L}
\rho_\alpha(L) P_\alpha^{\rho\sigma}(L)\nonumber\\
&&\qquad\times\Big[2L^2(R_\rho R_\sigma+P_\rho
P_\sigma)\;s_{_{R}}(P)s_{_{R}}(P+L)
s_{_{A}}(R)s_{_{A}}(R+L)\nonumber\\
&&\qquad\quad
-4(Q^2+2(P\cdot R-\overline{P}\cdot\overline{R})) R_\rho P_\sigma\;
s_{_{R}}(P)s_{_{R}}(P+L)
s_{_{A}}(R)s_{_{A}}(R+L)\nonumber\\
&&\qquad\quad-4 (Q^2+2(P\cdot R-\overline{P}\cdot\overline{R})) R_\rho R_\sigma \; 
s_{_{R}}(P)
[s_{_{A}}(R)]^2s_{_{A}}(R+L)\nonumber\\
&&\qquad\quad-4 (Q^2+2(P\cdot R-\overline{P}\cdot\overline{R})) P_\rho P_\sigma\;
s_{_{A}}(R)
[s_{_{R}}(P)]^2s_{_{R}}(P+L)
\Big]+{\rm c.c.}\; .\nonumber\\
&&
\label{eq:matrix-el}
\end{eqnarray}
Let us add that, since we are looking at photons whose virtuality is
not negligible, we have kept the terms in $Q^2$ when they have the
collinearly enhanced quark propagators.

\section{Integration over the quark momentum}
\subsection{Pole structure of the integrand and collinear enhancement}
We are now in a position to perform the integration over the quark
momentum. We perform this integration in the complex plane using the
theorem of residues, and we integrate first over the component of the
quark momentum which is parallel to the photon momentum\footnote{We
chose the photon momentum $Q$ to be along the $z$ axis. Therefore, by
definition, we have ${\imb q}_\perp=0$.}, i.e. the z-component.

For each term, we have propagators with a retarded prescription
multiplied by propagators with an advanced prescription. It is
convenient to perform in the complex plane the integration over the
longitudinal components of the quark momenta (the reference direction is
the photon momentum), by closing the contour around the upper
half-plane. In this approach, the collinear enhancement is due to the
fact that in a product like $s_{_{R}}(P)s_{_{A}}(R)$, there are pairs of
poles that are separated by a very small interval and are on opposite
sides of the real axis. As a consequence, the real axis is ``pinched''
by such a pair of poles, leading to a large contribution of the order of
the small separation between the poles.

Keeping only the contribution of the pinching pole, $p_z=p_0+({\imb
  p}_\perp^2+M_{\rm eff}^2)/2p_0$, the basic result we need is the
following (for $R=P+Q$)
\begin{equation}
\int_{-\infty}^{+\infty} {{dp_z}\over{2\pi}} s_{_{R}}(P)s_{_{A}}(R)F(p_z)
\approx
{1\over{2i q_0}} {1\over{{\imb p}_\perp^2+M_{\rm eff}^2-i{{p_0 r_0}\over{q_0}}\epsilon}}F(p_0)\; ,
\end{equation}
where, following \cite{AurenGKP2,AurenGKZ1}, we denote
\begin{equation}
M_{\rm eff}^2\equiv M_\infty^2+ {{Q^2}\over{q_0^2}} p_0 r_0\; .
\end{equation}
This result is valid if the function $F$ is not participating in the
collinear enhancement (i.e. is well behaved in the vicinity of the poles
of the quark propagator).  

\subsection{Contribution of the vertex diagram}
For the calculation of the first two terms in Eq.~(\ref{eq:matrix-el})
(i.e. the terms coming from the vertex correction diagram), we use
this method in order to perform the integral over the z-component of
the quark momentum in each loop, i.e. respectively $p_z$ and $p_z+l_z$
considered as independent variables. These terms become:
\begin{eqnarray}
&&\left.{\rm Im}\,\Pi_{_{R}}{}_\mu^\mu(Q)\right|_{\rm
vertex}\approx\nonumber\\
&&\qquad\approx 
{{e^2 g^2 N_c C_{_{F}}}\over{q_0^2}}
\int{{dp_0d^2{\imb
p}_\perp}\over{(2\pi)^3}}[n_{_{F}}(r_0)-n_{_{F}}(p_0)]
\int {{dl_0d^2{\imb l}_\perp}\over{(2\pi)^3}}{T\over{l_0}}
\nonumber\\
&&\qquad\times
\sum_{\alpha=T,L}
\rho_\alpha(L) P_\alpha^{\rho\sigma}(L)\nonumber\\
&&\qquad\times
\Big[L^2(R_\rho R_\sigma+P_\rho
P_\sigma)-2(Q^2+2(P\cdot R-\overline{P}\cdot\overline{R})) 
R_\rho P_\sigma\Big]
\Big|_{\empile{p_z=p_0}\over{l_z=l_0}}\nonumber\\
&&\qquad\times
{1\over{{\imb p}_\perp^2+M_{\rm eff}^2-i{{p_0 r_0}\over{q_0}}\epsilon}}\;
{1\over{({\imb p}_\perp+{\imb l}_\perp)^2+M_{\rm eff}^2-i{{p_0
r_0}\over{q_0}}\epsilon}}+\,{\rm c.c.}\; .
\end{eqnarray}
Next, we need to perform the contractions with the projectors. For any
pair of 4-vectors $A_\mu,B_\mu$ taken in the set $\{P_\mu,R_\mu\}$, we
have in the collinear limit:
\begin{equation}
A_\mu B_\nu P_{_{T}}^{\mu\nu}(L)=(\hat{\imb l}\cdot{\imb a})(\hat{\imb
l}\cdot{\imb b})-({\imb a}\cdot{\imb b})
\approx -A_\mu B_\nu P_{_{L}}^{\mu\nu}(L)\; ,
\end{equation}
where $\hat{\imb l}\equiv {\imb l}/l$. Noticing that all the transverse
components of ${\imb p}$ and ${\imb r}$ are much smaller than the
longitudinal components due to the collinear enhancement, we have also
\begin{equation}
\hat{\imb l}\cdot{\imb p}\approx p_0 {{l_0}\over{l}}\; ,
\qquad\hat{\imb l}\cdot{\imb r}\approx r_0 {{l_0}\over{l}}\; ,
\qquad{\imb p}\cdot{\imb r}\approx p_0 r_0\; ,
\end{equation}
so that
\begin{eqnarray}
&&P_\rho P_\sigma P_{_{T}}^{\rho\sigma}(L)
\approx -P_\rho P_\sigma P_{_{L}}^{\rho\sigma}(L)
\approx p_0^2 \Big({{l_0^2}\over{l^2}}-1\Big)\; ,\nonumber\\
&&R_\rho R_\sigma P_{_{T}}^{\rho\sigma}(L)
\approx -R_\rho R_\sigma P_{_{L}}^{\rho\sigma}(L)
\approx r_0^2 \Big({{l_0^2}\over{l^2}}-1\Big)\; ,\nonumber\\
&&P_\rho R_\sigma P_{_{T}}^{\rho\sigma}(L)
\approx -P_\rho R_\sigma P_{_{L}}^{\rho\sigma}(L)
\approx p_0 r_0 \Big({{l_0^2}\over{l^2}}-1\Big)\; .
\end{eqnarray}
It is convenient to use the variable $x\equiv l_0/l$ instead
of $l_0$ itself. However, since $l_z=l_0$ and $l^2=l_0^2+{\imb
l}_\perp^2$, the Jacobian of the transformation differs from $1$. A
simple calculation gives:
\begin{equation}
\Big(1-{{l_0^2}\over{l^2}}\Big){{dl_0}\over{l_0}}d(l_\perp^2)=
{{dx}\over x} d(l_\perp^2)\; .
\end{equation}
Therefore, this change of variable enables one to absorb all the
factors $1-l_0^2/l^2$ coming from the contraction with the
projectors\footnote{Since $L$ is space-like (as required for
  scattering processes like those of figure \ref{fig:processes}), the
  integration range for the variable $x$ is $[-1,+1]$, which can be
  replaced by $[0,1]$ using the parity of the integrand in the
  variable $x$.}.

Finally, taking $u\equiv p_\perp^2$ and $v\equiv l_\perp^2$ as the integration
variables, we can rewrite after carrying out the angular integrations:
\begin{eqnarray}
&&\!\!\!\!\left.{\rm Im}\,\Pi_{_{R}}{}_\mu^\mu(Q)\right|_{\rm
vertex}\approx 
-{{e^2 g^2 N_c C_{_{F}}}\over{32\pi^4}}{T\over{q_0^2}}
\int\limits_{-\infty}^{+\infty}dp_0[n_{_{F}}(r_0)-n_{_{F}}(p_0)]\nonumber\\
&&\quad\times
\int\limits_0^1{{dx}\over x}\int\limits_0^{+\infty} dv
\Big[(v+2M_\infty^2)(p_0^2+r_0^2)+2Q^2 p_0r_0\Big]\nonumber\\
&&\quad\times
\left[\!
{{2{\rm Im}\,\Pi_{_{T}}(x)}\over{(v+{\rm
Re}\,\Pi_{_{T}}(x))^2+({\rm Im}\,\Pi_{_{T}}(x))^2}}
\!-\!{{2{\rm Im}\,\Pi_{_{L}}(x)}\over{(v+{\rm
Re}\,\Pi_{_{L}}(x))^2+({\rm Im}\,\Pi_{_{L}}(x))^2}}\!
\right]
\nonumber\\
&&\quad\times\int\limits_0^{+\infty}du
{1\over{{u+\overline{M_{\rm eff}^2}}}}\;
{{{\rm sign}(u+v+M_{\rm
eff}^2)}\over{\left[(u+v+\overline{M_{\rm
eff}^2})^2-4uv\right]^{1/2}}}\,+{\rm c.c.}\; ,
\label{eq:expr-vertex}
\end{eqnarray}
where we denote
\begin{equation}
\overline{M_{\rm eff}^2}\equiv M_{\rm eff}^2-i {{p_0 r_0}\over{q_0}}\epsilon\; .
\end{equation}
This $i\epsilon$ prescription coming from the quark propagators will
turn out to be important when $M_{\rm eff}^2<0$.

\subsection{Contribution of the self-energy diagrams}
A similar strategy can be pursued for the contributions of the
self-energy diagrams, i.e. for the third and fourth terms in
Eq.~(\ref{eq:matrix-el}). Using the fact that these two terms are equal,
and performing the integration over $p_z$, we first obtain
\begin{eqnarray}
&&\left.{\rm Im}\,\Pi_{_{R}}{}_\mu^\mu(Q)\right|_{\rm
self}\nonumber\\
&&\qquad\qquad\approx 
{{4 e^2 g^2 N_c C_{_{F}}}\over{q_0^2}}
\int{{dp_0d^2{\imb
p}_\perp}\over{(2\pi)^3}}[n_{_{F}}(r_0)-n_{_{F}}(p_0)]
\int {{d^4L}\over{(2\pi)^4}}{T\over{l_0}}\nonumber\\
&&\qquad\qquad\times
\left[\rho_{_{L}}(L)-\rho_{_{T}}(L)\right] (Q^2+2(P\cdot R-\overline{P}\cdot\overline{R})) r^2 (1-\cos^2\theta_{rl})
\Big|_{p_z=p_0}\nonumber\\
&&\qquad\qquad\times
{1\over{\left[{\imb p}_\perp^2+M_{\rm eff}^2-i{{p_0
r_0}\over{q_0}}\epsilon\right]^2}}\; {{i p_0}\over{r_0l_0
-rl\cos\theta_{rl}-ir_0\epsilon}}+\,{\rm c.c.}\; ,\nonumber\\
&&
\end{eqnarray}
where $\theta_{rl}$ is the angle between the vectors ${\imb r}$ and
${\imb l}$. The integration over this angle is straightforward, since we
have:
\begin{eqnarray}
&&\int\limits_{-1}^{+1} {d\cos\theta_{rl}} {{1-\cos^2\theta_{rl}}\over{r_0l_0-ir_0\epsilon
-rl\cos\theta_{rl}}}\nonumber\\
&&\qquad\qquad={1\over {rl}}\left[1-{{(r_0l_0)^2}\over{(rl)^2}}
\right]
\ln\left({{r_0l_0-ir_0\epsilon+rl}\over{r_0l_0-ir_0\epsilon-rl}}\right)
+2{{r_0l_0}\over{(rl)^2}}\nonumber\\
&&\qquad\qquad\approx {1\over {rl}}\left[1-{{l_0^2}\over{l^2}}
\right]\,{\rm sign}(r_0)\left[\ln\left|{{l_0+l}\over{l_0-l}}\right|+i\pi\right]
+2{{r_0l_0}\over{(rl)^2}}\; .
\end{eqnarray}
Noticing that for soft $l_0$, the factor $\rho(L)T/l_0$ is an
even function of $l_0$, we can drop any term in the above result which
is odd in $l_0$. Therefore, only the imaginary part contributes. Using
this result, and introducing again the variables $x\equiv l_0/l$,
$u\equiv p_\perp^2$ and $v\equiv -L^2=l^2(1-x^2)$, we can write:
\begin{eqnarray}
&&\!\!\!\!\left.{\rm Im}\,\Pi_{_{R}}{}_\mu^\mu(Q)\right|_{\rm
self}\approx 
{{e^2 g^2 N_c C_{_{F}}}\over{32\pi^4}}{T\over{q_0^2}}
\int\limits_{-\infty}^{+\infty}dp_0[n_{_{F}}(r_0)-n_{_{F}}(p_0)]\nonumber\\
&&\quad\times
\int\limits_0^1{{dx}\over x}\int\limits_0^{+\infty} dv
\Big[2Q^2 p_0r_0+2M_\infty^2(p_0^2+r_0^2)\Big]\nonumber\\
&&\quad\times
\left[\!
{{2{\rm Im}\,\Pi_{_{T}}(x)}\over{(v+{\rm
Re}\,\Pi_{_{T}}(x))^2+({\rm Im}\,\Pi_{_{T}}(x))^2}}
\!-\!{{2{\rm Im}\,\Pi_{_{L}}(x)}\over{(v+{\rm
Re}\,\Pi_{_{L}}(x))^2+({\rm Im}\,\Pi_{_{L}}(x))^2}}\!
\right]
\nonumber\\
&&\quad\times\int\limits_0^{+\infty}du
{1\over{{\left[u+\overline{M_{\rm eff}^2}\right]^2}}}\;
\,+{\rm c.c.}\; ,
\label{eq:expr-self}
\end{eqnarray}

\subsection{Cancellation between vertex and self}
Combining Eqs.~(\ref{eq:expr-vertex}) and (\ref{eq:expr-self}), one
obtains
\begin{eqnarray}
&&\!\!\!\!{\rm Im}\,\Pi_{_{R}}{}_\mu^\mu(Q)\approx 
-{{e^2 g^2 N_c C_{_{F}}}\over{32\pi^4}}{T\over{q_0^2}}
\int\limits_{-\infty}^{+\infty}dp_0
[n_{_{F}}(r_0)-n_{_{F}}(p_0)]
\int\limits_0^1{{dx}\over x}\int\limits_0^{+\infty} dv\nonumber\\
&&\quad\times
\left[\!
{{2{\rm Im}\,\Pi_{_{T}}(x)}\over{(v+{\rm
Re}\,\Pi_{_{T}}(x))^2+({\rm Im}\,\Pi_{_{T}}(x))^2}}
\!-\!{{2{\rm Im}\,\Pi_{_{L}}(x)}\over{(v+{\rm
Re}\,\Pi_{_{L}}(x))^2+({\rm Im}\,\Pi_{_{L}}(x))^2}}\!
\right]\nonumber\\
&&\quad \times\int\limits_0^{+\infty}{{du}\over{u+\overline{M_{\rm eff}^2}}}\left\{v (p_0^2+r_0^2)  
{{{\rm sign}(u+v+M_{\rm
eff}^2)}\over{\left[(u+v+\overline{M_{\rm
eff}^2})^2-4uv\right]^{1/2}}}\right.\nonumber\\
&&\quad\left. + 2(Q^2 p_0r_0+M_\infty^2(p_0^2+r_0^2)) \left[
{{{\rm sign}(u+v+M_{\rm
eff}^2)}\over{\left[(u+v+\overline{M_{\rm
eff}^2})^2-4uv\right]^{1/2}}}
\!-\!
{1\over{u+\overline{M_{\rm eff}^2}}}
\right]\right\}\nonumber\\
&&\qquad\,+{\rm c.c.}\; .
\label{eq:ImPi}
\end{eqnarray}
Therefore, we see that there is a cancellation for the term in $Q^2$
between the contribution of the vertex and the contribution of the
self-energy corrections, in the limit where $v\to 0$, i.e. when the
momentum transferred by the gluon is small. This is an extension to
virtual photons of a well known cancellation in the case of real photons
\cite{LebedS1,LebedS2,CarriKP1,CarriK1,Bodek1,AurenGZ2,ArnolMY1,ArnolMY2}.
In particular, it ensures that the dilepton rate is not sensitive to the
gluon magnetic mass if the magnetic mass is small enough (say of order $g^2
T$).

\subsection{Integration over $p_\perp^2$}
At this stage, it is possible to perform analytically the integration
over the transverse momentum of the quark $u=p_\perp^2$. One has a
priori to distinguish two cases: $M_{\rm eff}^2>0$ and $M_{\rm
eff}^2<0$. The simplest case is $M_{\rm eff}^2>0$, for which the
denominators never vanish and for which the $i\epsilon$ prescription is
irrelevant (therefore, adding the complex conjugate just amounts to
multiply by a factor $2$). In this case, we have:
\begin{eqnarray}
&&\int\limits_0^{+\infty}\!\!
{{du}\over{u+\overline{M_{\rm eff}^2}}}{{{\rm sign}(u+v+M_{\rm
eff}^2)}\over{\left[(u+v+\overline{M_{\rm
eff}}^2)^2-4uv\right]^{1/2}}}+{\rm c.c.}\empile{=}\over{\epsilon\to 0^+}
{{8{\rm tanh}^{-1}\sqrt{{v\over{v+4M_{\rm eff}^2}}}}
\over{\sqrt{v(v+4M_{\rm eff}^2)}}}\; ,\nonumber\\
&&\int\limits_0^{+\infty}\!\!{{du}\over{\left[u+\overline{M_{\rm
eff}^2}\right]^2}}+{\rm c.c.}\empile{=}\over{\epsilon\to 0^+}
{2\over{M_{\rm eff}^2}}\; .
\label{eq:int-u-basic}
\end{eqnarray}

A similar calculation can be carried out when $M_{\rm eff}^2<0$. Note
that in this case the result of the integral over $u$ is a complex
number, but because we need to add its complex conjugate, only the real
part is important to us. In addition, the calculation shows that we have
to distinguish according to whether $v+4M_{\rm eff}^2>0$ or $v+4M_{\rm
eff}^2<0$. The results are the following (see the appendix
\ref{app:integral} for details):
\begin{eqnarray}
&&\bullet\quad{\rm If}\quad M_{\rm eff}^2<0\quad{\rm and}\quad
 v+4M_{\rm eff}^2>0\;:\nonumber\\
&&\!\!\!\!\int\limits_0^{+\infty}\!\!
{{du}\over{u+\overline{M_{\rm eff}^2}}}{{{\rm sign}(u+v+M_{\rm
eff}^2)}\over{\left[(u+v+\overline{M_{\rm
eff}^2})^2-4uv\right]^{1/2}}}+{\rm c.c.}\empile{=}\over{\epsilon\to 0^+}
{{8{\rm tanh}^{-1}\sqrt{{{v+4M_{\rm eff}^2}\over v}}}
\over{\sqrt{v(v+4M_{\rm eff}^2)}}}\; ,\nonumber\\
&&\bullet\quad{\rm If}\quad M_{\rm eff}^2<0\quad{\rm and}\quad
 v+4M_{\rm eff}^2<0\;:\nonumber\\
&&\!\!\!\!\int\limits_0^{+\infty}\!\!
{{du}\over{u+\overline{M_{\rm eff}^2}}}{{{\rm sign}(u+v+M_{\rm
eff}^2)}\over{\left[(u+v+\overline{M_{\rm
eff}^2})^2-4uv\right]^{1/2}}}+{\rm c.c.}\empile{=}\over{\epsilon\to 0^+}-
{{8{\rm tan}^{-1}\sqrt{-{v\over{v+4M_{\rm eff}^2}}}}
\over{\sqrt{-v(v+4M_{\rm eff}^2)}}}\; .\nonumber\\
&&
\label{eq:A-neg}
\end{eqnarray}
One notices that those results could have been obtained easily by
taking the real part of the analytic continuation of the $M_{\rm
  eff}^2>0$ result. Indeed, if $M_{\rm eff}^2<0$ and $v+4M_{\rm
  eff}^2>0$, then $\sqrt{v/(v+4M_{\rm eff}^2)}>1$ and this analytic
continuation just amounts to replace the argument of the ${\rm
  tanh}^{-1}$ in Eq.~(\ref{eq:int-u-basic}) by its inverse. If on the
contrary $v+4M_{\rm eff}^2<0$ then the square root in
Eq.~(\ref{eq:int-u-basic}) is a purely imaginary number, and by using
$iX{\rm tanh}^{-1}(iX)=-X{\rm tan}^{-1}(X)$ one obtains easily the
correct answer.

Denoting $v\equiv M_{\rm eff}^2w$ and introducing the following set of
functions\footnote{Or their analytic continuations if $M_{\rm eff}^2$ is
negative.}:
\begin{eqnarray}
&&\!\!\!\!\!J_{_{T,L}}\equiv M_{\rm eff}^2 \int\limits_{0}^{1}\! {{dx}\over{x}} {\rm
Im}\,\Pi_{_{T,L}}(x) \!\int\limits_{0}^{+\infty}\!\! dw
{{\sqrt{w/(w+4)} {\rm tanh}^{-1}\sqrt{w/(w+4)}}\over
{(M_{\rm eff}^2w+{\rm Re}\,\Pi_{_{T,L}}(x))^2+({\rm Im}\,\Pi_{_{T,L}}(x))^2}}\;
,\nonumber\\
&&\!\!\!\!\!K_{_{T,L}}\equiv M_{\rm eff}^2 \int\limits_{0}^{1}\! {{dx}\over{x}} {\rm
Im}\,\Pi_{_{T,L}}(x) \!\int\limits_{0}^{+\infty}\!\! {{dw}\over w}
{{\sqrt{w/(w+4)} {\rm tanh}^{-1}\sqrt{w/(w+4)}-w/4}\over
{(M_{\rm eff}^2w+{\rm Re}\,\Pi_{_{T,L}}(x))^2+({\rm Im}\,\Pi_{_{T,L}}(x))^2}}\;
,\nonumber\\
&&
\label{eq:JT}
\end{eqnarray}
one can write a very compact expression for the 2-loop photon
polarization tensor:
\begin{eqnarray}
&&\!\!\!\!{\rm Im}\,\Pi_{_{R}}{}_\mu^\mu(Q)\approx 
-{{e^2 g^2 N_c C_{_{F}}}\over{2\pi^4}}{T\over{q_0^2}}
\int\limits_{-\infty}^{+\infty}dp_0[n_{_{F}}(r_0)-n_{_{F}}(p_0)]\nonumber\\
&&\qquad\times\left\{(p_0^2+r_0^2)(J_{_{T}}-J_{_{L}})+2{{Q^2 p_0r_0+M_\infty^2(p_0^2+r_0^2)}\over{M_{\rm
eff}^2}} (K_{_{T}}-K_{_{L}})\right\}\; .\nonumber\\
&&
\label{eq:rate-final}
\end{eqnarray}
Note that the above defined $J_{_{T,L}}$ match those already defined in
\cite{AurenGKP2}. Having this in mind, one can recover the limit of real
photons.  Indeed, in the limit where $Q^2\to 0$ the quantities
$J_{_{T,L}}$ and $K_{_{T,L}}$ become independent of the quark energy
$p_0$, and can be factored out of the integral (they depend only on
$M_\infty/m_{\rm g}$). Therefore, we obtain:
\begin{eqnarray}
\!\lim_{Q^2=0} {\rm Im}\,\Pi_{_{R}}{}_\mu^\mu(Q)&\approx& {{e^2 g^2
N_c
C_{_{F}}}\over{2\pi^4}}{{T}\over{q_0^2}}(J_{_{T}}-J_{_{L}}+2K_{_{T}}-2K_{_{L}})\nonumber\\
&&\qquad\times\int\limits_{-\infty}^{+\infty}\!\!dp_0[n_{_{F}}(r_0)-n_{_{F}}(p_0)](p_0^2+r_0^2)\nonumber\\
&=&{{e^2 g^2N_c C_{_{F}}}\over{3\pi^4}}(J_{_{T}}-J_{_{L}}+2K_{_{T}}-2K_{_{L}})
\left[q_0T+\pi^2{{T^3}\over{q_0}}\right]\; .\nonumber\\
&&
\end{eqnarray}
Note that the integration over $p_0$ is performed exactly
here\footnote{Indeed, one can prove:
\begin{equation}
\int_{-\infty}^{+\infty}dx\left[{1\over{e^x+1}}-{1\over{e^{x+y}+1}}\right]
\left[x^2+(x+y)^2\right]={2\over 3}\left[y^3+\pi^2 y\right]\; .
\end{equation}}.  In
the papers \cite{AurenGKP2,AurenGKZ1}, the numerical evaluation of
$J_{_{T,L}}$ was overestimated by a factor of $4$, as first pointed
out in \cite{Mohan1,Mohan2} and independently in \cite{SteffT1}. We
also see that taking correctly into account the HTL correction to the
vertices, which was ignored in \cite{AurenGKP2,AurenGKZ1}, brings
another term of the same order, namely $2K_{_{T}}-2K_{_{L}}$.
Numerically, this term is a $30\%$ negative correction for $N_c=3$ and
$N_{_{F}}=2$ or $3$. Although the references \cite{ArnolMY1,ArnolMY2} did
not consider this HTL vertex correction, they obtained the correct
answer\footnote{But we do not understand how they could find an
  agreement with our incomplete result of \cite{AurenGKZ1}, as stated
  on page 8 of \cite{ArnolMY2}.} for real photons\footnote{Their
  result is incomplete for virtual photons, because they did not
  include the longitudinal mode of the massive photon.}  because they
chose to calculate only the transverse part of the polarization tensor
($\Pi_{11}+\Pi_{22}$). Indeed, it is easy to check that the HTL
correction to the vertex modifies only the component $\Pi_{zz}$.

One can also add that Eq.~(\ref{eq:rate-final}) combines in a single
integral the contributions of all the processes in figure
\ref{fig:processes}. The bremsstrahlung is obtained for $p_0>0$
(bremsstrahlung of a quark) and $p_0<-q_0$ (bremsstrahlung of an
antiquark), while for $-q_0<p_0<0$ one gets the off-shell
annihilation.

\subsection{Exact integration over $x$}
The advantage of the expression given in Eq.~(\ref{eq:rate-final}) is
that the integrals $J_{_{T,L}}$ and $K_{_{T,L}}$ it contains are
functions of $M_{\rm eff}^2/m_{\rm g}^2$ which can be studied rather simply
analytically. Indeed, we show in a separate paper \cite{AurenGZ4} that
the result of the integration over the variable $x$ is in fact extremely
simple thanks to the use of sum rules. We show that for a general enough
self-energy (see \cite{AurenGZ4} for the conditions under which it is
true), one has:
\begin{equation}
\int_{0}^{1}{{dx}\over{x}} 
{{2{\rm Im}\,\Pi(x)}\over{(z+{\rm Re}\,\Pi(x))^2+({\rm
Im}\,\Pi(x))^2}}=\pi\left[{1\over{z+{\rm Re}\,\Pi(\infty)}}
-
{1\over{z+{\rm Re}\,\Pi(0)}} \right]\, .
\label{eq:fz}
\end{equation}
From there, it is possible to give integral expressions for the
functions $J_{_{T,L}}$ and $K_{_{T,L}}$ that are much simpler than
Eq.~(\ref{eq:JT}) and are very well suited for the derivation of various
asymptotic expressions. In this paper, we just quote the results we need
without proof, and refer the reader to \cite{AurenGZ4} for a thorough
justification. For instance, we obtain:
\begin{eqnarray}
&&J_{_{L}}-J_{_{T}}=\pi \int\limits_0^1 du 
{{{\rm tanh}^{-1}(u)}\over{({{4M_{\rm eff}^2}\over{3 m_{\rm g}^2}}-1)u^2+1}}\; ,\nonumber\\
&&K_{_{L}}-K_{_{T}}=\pi\left[
{1\over 4}+{1\over 8}\ln\left(
{{M_{\rm eff}^2}\over{3 m_{\rm g}^2}}
\right)-{{M_{\rm eff}^2}\over{3 m_{\rm g}^2}}
\int\limits_0^1 du 
{{{\rm tanh}^{-1}(u)}\over{({{4M_{\rm eff}^2}\over{3 m_{\rm g}^2}}-1)u^2+1}}
\right]\; .
\end{eqnarray}

For real photons ($Q^2=0$, $M_{\rm eff}^2=M_\infty^2$) and for $N_c=3$
colors, we have:
\begin{equation}
{{4M_{\rm eff}^2}\over{3m_{\rm g}^2}}={8\over{6+N_{_{F}}}}\; .
\end{equation}
Therefore, there is an accidental simplification for $N_{_{F}}=2$
flavors, as this ratio is then equal to $1$. In that case, we have
\begin{eqnarray}
&&J_{_{L}}-J_{_{T}}\Big|_{N_c=3,N_{_{F}}=2}=\pi\ln(2)\; ,\nonumber\\
&&K_{_{L}}-K_{_{T}}\Big|_{N_c=3,N_{_{F}}=2}={\pi\over 4}(1-2\ln(2))\; ,
\end{eqnarray}
and therefore the 4-dimensional integral of Eq.~(\ref{eq:ImPi}) can be
done exactly:
\begin{equation}
\lim_{Q^2=0} {\rm Im}\,\Pi_{_{R}}{}_\mu^\mu(Q)\Big|_{{N_c=3}\atop{N_{_{F}}=2}}=-{{2e^2 g^2}\over{3\pi^3}}
\left[q_0T+\pi^2{{T^3}\over{q_0}}\right]\; .
\end{equation}
Note that this expression is valid for hypothetical quarks of electrical
charge $e$. If the two quark species under consideration are $u$ and $d$
quarks, it must be multiplied by a factor $(1/3)^2+(2/3)^2=5/9$.

\section{Behavior near the tree-level threshold}
\subsection{Analytic expressions near $Q^2=4M^2_\infty$}
A close look at Eq.~(\ref{eq:rate-final}) seem to indicate that there
could be problems due to the denominator $M_{\rm eff}^2$ for the second term  since this effective mass parameter can vanish if $Q^2 \ge
4M_\infty^2$. This is not a problem as long as the zeros of $M_{\rm
eff}^2$ (in the variable $p_0$) are simple zeros, since this term should
be understood with a principal value prescription. This is the generic
case, since the zeros are simple for any $Q^2$ not equal to
$4M_\infty^2$.  However, there is a problem for a photon invariant mass
$Q^2=4M_\infty^2$, i.e. at the threshold for the tree level process
$q\bar{q}\to\gamma^*$. Indeed, for this value of $Q^2$, the quantity
$M_{\rm eff}^2$ has a double pole at $p_0=-q_0/2$. Being a double pole,
it cannot be dealt with a principal part prescription and makes the
result infinite.

In order to make this statement more precise, we derive here the
analytic behavior of ${\rm Im}\,\Pi_{_{R}}{}_\mu^\mu(Q)$ for $Q^2$ in
the vicinity of $4M_\infty^2$. This calculation is made simple by the
fact that we want to extract only the diverging pieces near
$Q^2=4M_\infty^2$. We can therefore drop the term in $J_{_{T,L}}$ since
it is finite, and replace $p_0 r_0 Q^2$ by $-q_0^2 M_\infty^2$ (since
$Q^2\approx 4M_\infty^2$ and $p_0\approx -q_0/2$ is the location of the
singularity), which gives
\begin{eqnarray}
\!\!\!\!{\rm Im}\,\Pi_{_{R}}{}_\mu^\mu(Q)\empile{\approx}\over{\rm threshold} 
{{e^2 g^2 N_c C_{_{F}}}\over{2\pi^4}}{TM_\infty^2}
\int\limits_{-\infty}^{+\infty}dp_0[n_{_{F}}(r_0)-n_{_{F}}(p_0)]
{{K_{_{T}}-K_{_{L}}}\over{M_{\rm eff}^2}}\; .
\end{eqnarray} 
In order to simplify further the calculation, let us limit ourselves to
a large photon energy $q_0\gg T$. In this case, one can check that the
statistical weight $n_{_{F}}(p_0)-n_{_{F}}(p_0+q_0)$ can be approximated
by a function which is $1$ in the range $[-q_0,0]$ and $0$
elsewhere. Doing so is accurate up to terms suppressed by at least one
power of $T/q_0\ll 1$. Due to a symmetry of the integrand, we can in
fact integrate only over $[-q_0/2,0]$ and multiply the result by a
factor $2$.

\subsection{Case $Q^2<4M_\infty^2$}
At this point, we have to distinguish two cases, depending on whether
$Q^2<4M_\infty^2$ or $Q^2>4M_\infty^2$. Let us start with the case
$Q^2<4M_\infty^2$. In this case, it is convenient to replace the
integration variable $p_0$ by a variable $z$ defined as follows
\begin{equation}
p_0+{{q_0}\over{2}}\equiv {{q_0
M_\infty}\over{\sqrt{Q^2}}}\sqrt{1-{{Q^2}\over{4M_\infty^2}}}\; z\; .
\end{equation}
The range $p_0\in[-q_0/2,0]$ is mapped onto
$z\in[0,\sqrt{Q^2/(4M_\infty^2-Q^2)}]$ in this
transformation\footnote{Note that the upper bound for $z$ is large near
the threshold, and can be replaced by $+\infty$ in this calculation.},
and the expression of $M_{\rm eff}^2$ becomes:
\begin{equation}
M_{\rm eff}^2={1\over 4}(4M_\infty^2-Q^2)(z^2+1)\; .
\end{equation}
This leads to the following expression for the photon polarization
tensor near the threshold:
\begin{eqnarray}
\!\!\!\!{\rm Im}\,\Pi_{_{R}}{}_\mu^\mu(Q)\empile{\approx}\over{\rm
Q^2\approx (4M_\infty^2)^-} 
-{{2e^2 g^2 N_c C_{_{F}}}\over{\pi^4}}
{{q_0 T M_\infty^2}\over{\sqrt{Q^2(4M_\infty^2-Q^2)}}}
\int\limits_{0}^{+\infty}dz\,
{{K_{_{T}}-K_{_{L}}}\over{z^2+1}}\; .
\end{eqnarray} 
At this point, we need only to know the behavior at small positive
$M_{\rm eff}^2$ of the functions $K_{_{T,L}}$. In \cite{AurenGZ4}, we
prove the following results:
\begin{eqnarray}
&&K_{_{T}}\empile{\approx}\over{M_{\rm eff}\ll m_{\rm g}} {\pi\over 8}
\left[\ln\Big({{m_{\rm g}^2}\over{M_{\rm eff}^2}}\Big)-2\right]\; ,\nonumber\\
&&K_{_{L}}\empile{\approx}\over{M_{\rm eff}\ll m_{\rm g}}-{{\pi\ln(3)}\over
8}\; ,
\label{eq:K-asympt}
\end{eqnarray}
where the terms neglected vanish when $M_{\rm eff}^2\to 0$ and therefore
do not contribute to the singular behavior near the threshold.  The
longitudinal contribution to this self-energy is therefore given by
\begin{equation}
{\rm Im}\,\Pi_{_{R}}{}_\mu^\mu(Q)\Big|_{_{L}}\empile{\approx}\over{\rm
Q^2\approx (4M_\infty^2)^-} 
-{{e^2 g^2 N_c C_{_{F}}}\over{8\pi^2}}
{{q_0 T M_\infty^2 \ln(3)}\over{\sqrt{Q^2(4M_\infty^2-Q^2)}}}
\; ,
\label{eq:below-th-L}
\end{equation}
while the transverse contribution is\footnote{For this calculation, we
need the following definite integral:
\begin{equation}
\int_0^{+\infty}dz {{\ln(z^2+1)}\over{z^2+1}}=\pi\ln(2)\; .
\end{equation}}
\begin{equation}
{\rm Im}\,\Pi_{_{R}}{}_\mu^\mu(Q)\Big|_{_{T}}\!\empile{\approx}\over{\rm
Q^2\approx (4M_\infty^2)^-}\! \!\!
-{{e^2 g^2 N_c C_{_{F}}}\over{8\pi^2}}\!
{{q_0 T M_\infty^2}\over{\sqrt{Q^2(4M_\infty^2-Q^2)}}}\!\left[
\ln\left(\!{{m^2_{\rm g}}\over{4M_\infty^2-Q^2}}\!\right)-2
\right]
\, .
\label{eq:below-th-T}
\end{equation}
Note that these terms arise only from the cuts $(b)$ and $(b^\prime)$
of figure \ref{fig:diagrams} (the cuts $(a)$ and $(c)$ are zero in
this domain of $Q^2$).

\subsection{Case $Q^2>4M_\infty$}
If the invariant mass squared is above the threshold, we need to make a
different change of variables, because $M_{\rm eff}^2$ has zeros in the
integration range. Now, the appropriate change of variables is
\begin{equation}
p_0+{{q_0}\over{2}}\equiv {{q_0
M_\infty}\over{\sqrt{Q^2}}}\sqrt{{{Q^2}\over{4M_\infty^2}}-1}\; z\; ,
\end{equation}
which leads to
\begin{equation}
M_{\rm eff}^2={1\over 4}(Q^2-4M_\infty^2)(z^2-1)\; .
\end{equation}
Again, when we are very close to the threshold, the range $[-q_0/2,0]$
is mapped on $z\in[0,+\infty[$, which enables to write the following
expression for the photon polarization tensor above the threshold:
\begin{eqnarray}
\!\!\!\!{\rm Im}\,\Pi_{_{R}}{}_\mu^\mu(Q)\empile{\approx}\over{\rm
Q^2\approx (4M_\infty^2)^+} 
-{{2e^2 g^2 N_c C_{_{F}}}\over{\pi^4}}
{{q_0 T M_\infty^2}\over{\sqrt{Q^2(Q^2-4M_\infty^2)}}}
\int\limits_{0}^{+\infty}dz\,
{{K_{_{T}}-K_{_{L}}}\over{z^2-1}}\; .
\end{eqnarray}
The pole at $z=1$ should be handled with a principal value
prescription. We can make use again of Eqs.~(\ref{eq:K-asympt}) (with
the difference that the absolute value of $M_{\rm eff}^2$ should enter
in the logarithm for $K_{_{T}}$), and obtain for the longitudinal
contribution:
\begin{equation}
{\rm Im}\,\Pi_{_{R}}{}_\mu^\mu(Q)\Big|_{_{L}}\empile{\approx}\over{\rm
Q^2\approx (4M_\infty^2)^+} \!\!
-{{e^2 g^2 N_c C_{_{F}}}\over{4\pi^3}}
{{q_0 T M_\infty^2 \ln(3)}\over{\sqrt{Q^2(Q^2-4M_\infty^2)}}}
\int_0^{+\infty}{{dz}\over{z^2-1}}=0\; .
\end{equation}
However, the integral over $z$ is vanishing due to the principal part
prescription. In other words, the longitudinal contribution has no
singular part above the threshold. For the same reason, the non
vanishing singular piece in the transverse contribution is
\begin{equation}
{\rm Im}\,\Pi_{_{R}}{}_\mu^\mu(Q)\Big|_{_{T}}\empile{\approx}\over{\rm
Q^2\approx (4M_\infty^2)^+} 
-{{e^2 g^2 N_c C_{_{F}}}\over{4\pi^3}}
{{q_0 T M_\infty^2}\over{\sqrt{Q^2(Q^2-4M_\infty^2)}}}\int_0^{+\infty}dz\,{{\ln|1-z^2|}\over{1-z^2}}
\; .
\end{equation}
In order to compute the final integral, it is convenient to split it in
two terms $z\in[0,1]$ and $z\in[1,+\infty[$, and to perform the change
of variables $z\to 1/z$ in the second term. Upon applying this
transformation, the integral over $z$ becomes
\begin{equation}
\int_0^{+\infty}dz\,{{\ln|1-z^2|}\over{1-z^2}}=2\int_0^1 dz\,
{{\ln(z)}\over{1-z^2}}=-{{\pi^2}\over{4}}\; .
\end{equation}
Therefore, the singular part of the transverse contribution above the
threshold reads
\begin{equation}
{\rm Im}\,\Pi_{_{R}}{}_\mu^\mu(Q)\Big|_{_{T}}\empile{\approx}\over{\rm
Q^2\approx (4M_\infty^2)^+} 
{{e^2 g^2 N_c C_{_{F}}}\over{16\pi}}
{{q_0 T M_\infty^2}\over{\sqrt{Q^2(Q^2-4M_\infty^2)}}}
\; .
\label{eq:above-th-T}
\end{equation}
The slightly less singular behavior obtained for $Q^2>4M_\infty^2$ is
due to a partial cancellation between the cuts $(a),(c)$ and
$(b),(b^\prime)$. One can also note that the singularities exhibited in
this section are integrable: the energy spectrum, integrated over the
photon mass, is therefore always finite.

\subsection{Numerical illustration}
The above formulas for the singular behavior of the photon rate near
$Q^2=4M_\infty^2$ can be checked by a full numerical evaluation of
Eq.~(\ref{eq:rate-final}). The numerical evaluation is done for $N_c=3$
colors and $N_f=2$ light flavors, with a strong coupling constant $g=2$
(i.e. $\alpha_{_{S}}\approx 0.32$) and a photon energy $q_0/T=50$.
\begin{figure}[htb]
\centerline{\resizebox*{!}{7cm}{\rotatebox{-90}{\includegraphics{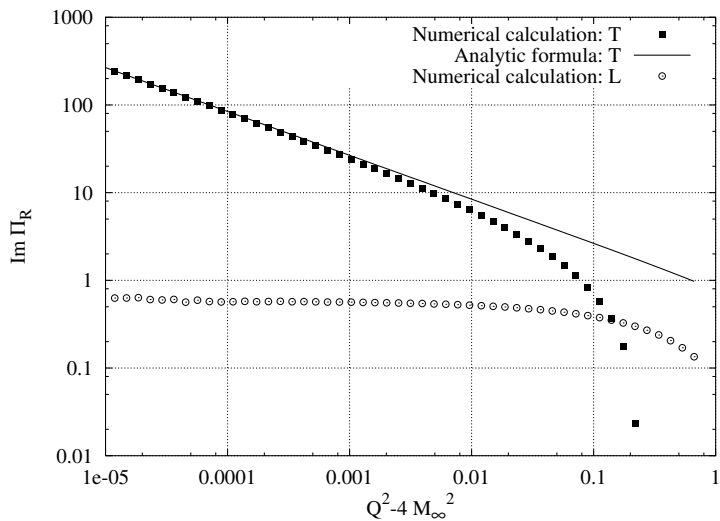}}}}
\caption{\label{fig:above-th} Behavior of the transverse and
longitudinal 2-loop self-energy just above the tree-level threshold. The
behavior of the transverse contribution is compared to the analytic
formula Eq.~(\ref{eq:above-th-T}) for the singular terms. Dimensionful
quantities are in units of the temperature $T$.}
\end{figure}
In figure \ref{fig:above-th}, one can see clearly that there is no
divergence at $Q^2=4M_\infty^2$ in the longitudinal contribution, and
that there is a divergence in the transverse contribution, correctly
reproduced by Eq.~(\ref{eq:above-th-T}).

\begin{figure}[htb]
\centerline{\resizebox*{!}{7cm}{\rotatebox{-90}{\includegraphics{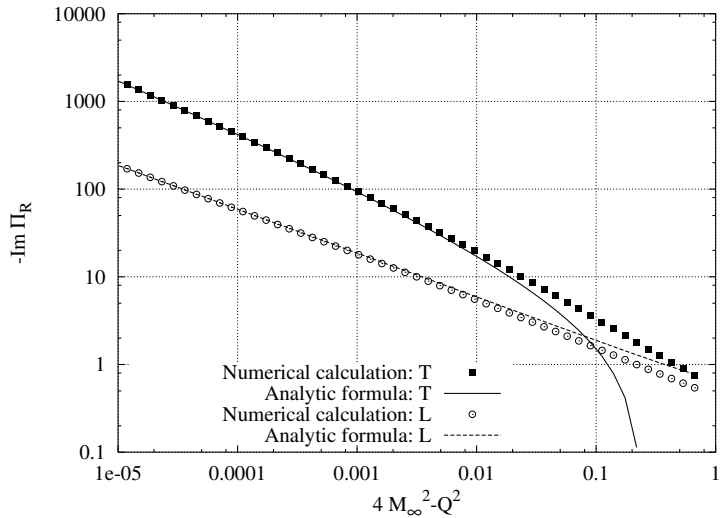}}}}
\caption{\label{fig:below-th} Behavior of the transverse and
longitudinal 2-loop self-energy just below the tree-level threshold. The
numerical results are compared to the analytic formulas
Eqs.~(\ref{eq:below-th-T}) and (\ref{eq:below-th-L}) for the singular
terms. Dimensionful quantities are in units of the temperature $T$. Note
that we display $-{\rm Im}\,\Pi_{_{R}}$ and not ${\rm Im}\,\Pi_{_{R}}$.}
\end{figure}

Figure \ref{fig:below-th} presents the same results below the
threshold. Here, both the transverse and longitudinal
contributions are singular, in very good agreement with the analytic
prediction of Eqs.~(\ref{eq:below-th-T}) and (\ref{eq:below-th-L}).

We observe that the 2-loop photon polarization tensor considered in this
paper is not defined if $Q^2=4M_\infty^2$, and more generally that the
perturbative expansion leads to inconsistent results in the vicinity of
the tree level threshold. This could have been expected on very general
grounds \cite{CatanW1}. Indeed, it is well known in perturbation theory
that if a contribution at order $n$ has a phase-space constraint (like
the $\theta(Q^2-4M_\infty^2)$ for $q\bar{q}\to\gamma^*$ at tree level),
then higher order corrections to this contribution exhibit a singularity
at the point where the constraint starts. A correct assessment of the
behavior of this process near the phase-space boundary requires usually
the resummation of an infinite number of terms.

In our case, among the next order corrections are a correction $\delta
M_\infty^2$ to the thermal mass of the hard quark, as well as a width
$\Gamma$ for the quark. Particularly important is the width which must
be resummed whenever $M_{\rm eff}^2$ is small
\cite{AurenGZ2,ArnolMY1,ArnolMY2} (and this threshold problem is due
to the possibility that $M_{\rm eff}^2$ vanishes). However, gauge
invariance dictates that ladder corrections to the $q\bar{q}\gamma$
vertex be also resummed. Therefore, one can anticipate that a complete
treatment near the threshold involves a simultaneous resummation of a
width on the quark propagator, and of the ladder corrections to the
vertex where the photon is attached. Since virtual photons have a
physical longitudinal mode, doing this requires to extend the work of
\cite{ArnolMY1,ArnolMY2} in order to resum also the LPM corrections
for the longitudinal mode \cite{WorkIP1}.

\subsection{Positiveness of the rate}
\begin{figure}[htb]
\centerline{\resizebox*{!}{7cm}{\rotatebox{-90}{\includegraphics{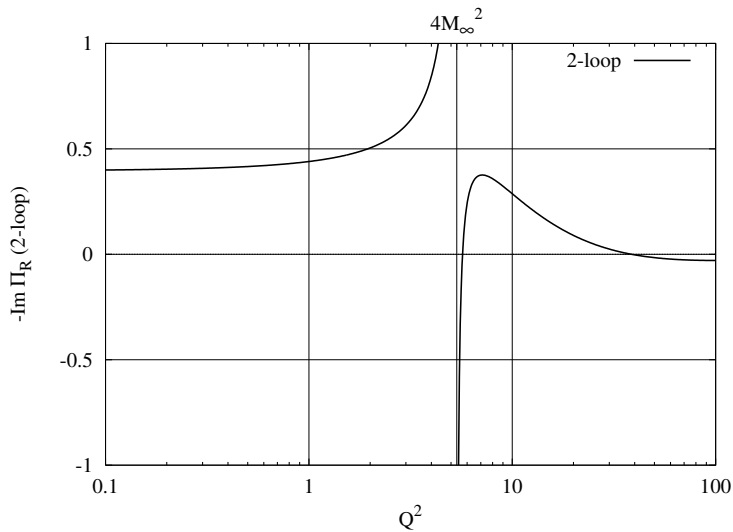}}}}
\caption{\label{fig:2-loop} Sum of the transverse and longitudinal 
  contributions to the 2-loop photon polarization tensor. The photon
  energy is set to $q_0/T=50$. We see clearly that there is a region
  of $Q^2$ where $-{\rm Im}\,\Pi_{_{R}}(Q)$ is negative. Dimensionful
  quantities are in units of $T$.}
\end{figure}

Away from the threshold, the 2-loop photon polarization tensor if
finite, but may have the wrong sign in order to give a positive
dilepton production rate, as illustrated in figure \ref{fig:2-loop}.
We see that this self-energy would lead to a negative contribution to
the photon rate immediately above $Q^2=4M_\infty^2$. This by itself is
not enough to indicate a violation of unitarity since above the
threshold the 2-loop contributions are nothing but {\sl higher order}
contributions to the tree-level process. They may
be negative, but are suppressed by a power of $\alpha_{_{S}}$, so that
the total rate should remain positive.

\begin{figure}[htb]
\centerline{\resizebox*{!}{7cm}{\rotatebox{-90}{\includegraphics{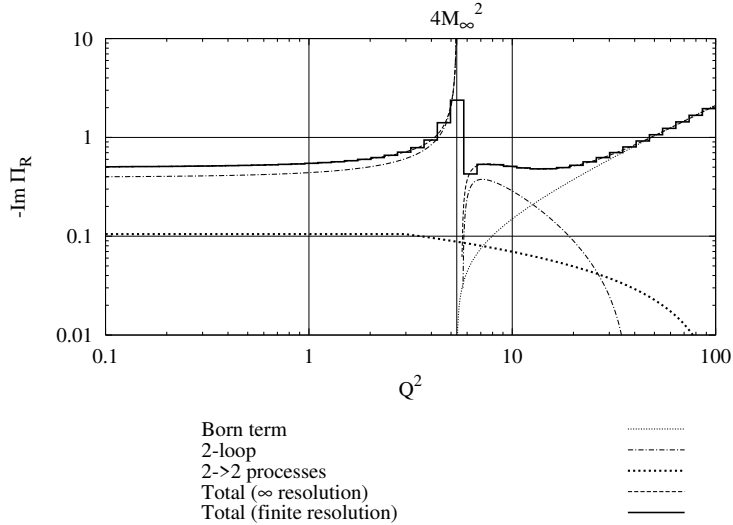}}}}
\caption{\label{fig:born-2loop} Born term, $2\to 2$ processes, and
  2-loop contributions. Also shown is the total yield (up to 2-loop)
  with a finite mass resolution. The photon energy is set to
  $q0/T=50$. Dimensionful quantities are in units of $T$.}
\end{figure}

Let us illustrate more graphically this point. The 1-loop diagram
contains only the direct production of a virtual photon by the
annihilation of a quark and an antiquark
\setbox1=\hbox to 1.7cm{{\resizebox*{1.7cm}{!}{\includegraphics{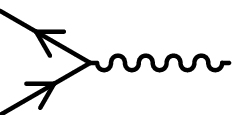}}}}
\begin{eqnarray}
\Bigg|\,\raise -3mm\box1\,\Bigg|^2\; .\nonumber
\end{eqnarray}
However, when we consider the following two-loop diagrams,
we have not only new processes like bremsstrahlung and off-shell annihilation
\setbox1=\hbox to 3.2cm{{\resizebox*{3.2cm}{!}{\includegraphics{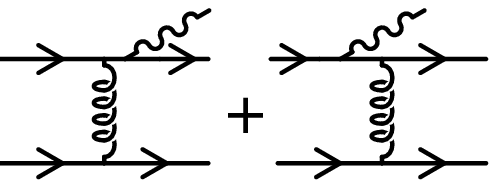}}}}
\setbox2=\hbox to 3.2cm{{\resizebox*{3.2cm}{!}{\includegraphics{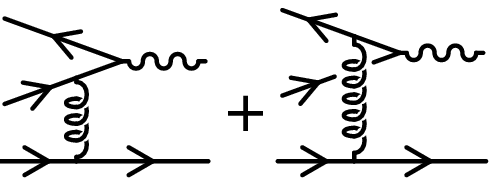}}}}
\begin{eqnarray}
\Bigg|\,\raise -5mm\box1\,\Bigg|^2+\Bigg|\,\raise -5mm\box2\,\Bigg|^2\nonumber
\end{eqnarray}
but also the interference between the Born level amplitude and loop
corrections to them
\setbox1=\hbox to 1.7cm{{\resizebox*{1.7cm}{!}{\includegraphics{process_born.ps}}}}
\setbox2=\hbox to 5cm{\hfil{\resizebox*{5cm}{!}{\includegraphics{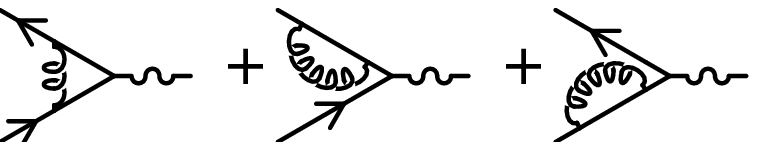}}}}
\begin{eqnarray}
\Bigg[\,\raise -3mm\box1\,\Bigg]\Bigg[\,\raise -3.5mm\box2\,\Bigg]^*+{\rm c.c.}\; .\nonumber
\end{eqnarray}
Naturally, these interference terms can be negative, and are
responsible for the fact that the 2-loop contribution is sometimes
negative. Therefore, even if we add up the tree level contribution
\begin{equation}
{\rm Im}\,\Pi_{_{R}}{}_\mu^\mu(Q)\Big|_{\rm Born}\approx -{{e^2 N_c}\over{4\pi}} \sqrt{Q^2(Q^2-4M_\infty^2)}\; ,
\end{equation}
it may happen that we get a negative rate
in the vicinity of the tree-level threshold, due to the breakdown of
perturbation theory at this particular point. This is in fact what
happens, as shown in figure \ref{fig:born-2loop}.
We show on this plot that the sum of the Born term and the 2-loop
terms has the wrong sign just above $Q^2=4M_\infty^2$. This indicates
that the 2-loop result should not be trusted in this region. The solid
curve shows that by dividing the $Q^2$ range in finite size bins (as
would be the case in any realistic experimental situation, where there
is only a limited mass resolution), one may average out positive and
negative contributions, and get a rate that is always positive. This
observation should not be considered as the definitive cure of this
problem, which should consist in a resummation to all orders. Also
shown on this plot for the sake of completeness is the contribution of
the processes $q\bar{q}\to g\gamma^*$ and $qg\to q\gamma^*$,
calculated in \cite{AltheR1}. These processes appear in the 1-loop HTL
photon polarization tensor, and contribute also at the order $e^2g^2$.
The corresponding contribution to ${\rm Im}\,\Pi_{_{R}}{}^\mu_\mu$ is
given by:
\begin{equation}
{\rm Im}\,\Pi_{_{R}}{}_\mu^\mu(Q)\Big|_{\rm 2\to 2}\approx -{{e^2 g^2
N_c C_{_{F}}}\over{16\pi}} T^2 \left[ \ln\left({{2q_0T}\over{Q^2}}\right)
+1+{{\ln(2)}\over{3}}-\gamma+{{\zeta^\prime(2)}\over{\zeta(2)}}\right]\; .
\label{eq:2to2} 
\end{equation}
Note that this expression should not be extrapolated to very small
photon masses. At small $Q^2$, the $Q^2$ in the logarithm is eventually
replaced by the quark thermal mass \cite{BaierNNR1,KapusLS1}. In
order to take simply this effect into account, we limit the growth of
Eq.~(\ref{eq:2to2}) at small $Q^2$ by replacing it by the $Q^2=0$ result
of \cite{BaierNNR1,KapusLS1} whenever Eq.~(\ref{eq:2to2}) would give a
larger result. It would be interesting to recalculate these $2\to 2$
processes for photon masses comparable to thermal masses in order to
obtain a more correct matching between Eq.~(\ref{eq:2to2}) and the
$Q^2=0$ limit.

\section{Phenomenology}
In this section, we present some results in a less academic
situation. We choose parameters as they may appear for heavy ion
collisions at LHC. The temperature is set to $1$ GeV\footnote{This may
be on the high side, but most of the photons and light dileptons are
produced in the early stages of the plasma evolution, when the
temperature is still very high.}. We take a coupling constant $g=2$,
i.e. $\alpha_{_{S}}\approx 0.32$. The number of colors is set to $N_c=3$
and we take two flavors ($u$ and $d$, with respective electric charges
$2/3$ and $-1/3$).

\subsection{Mass spectrum}
\begin{figure}
\centerline{\resizebox*{!}{7cm}{\rotatebox{-90}{\includegraphics{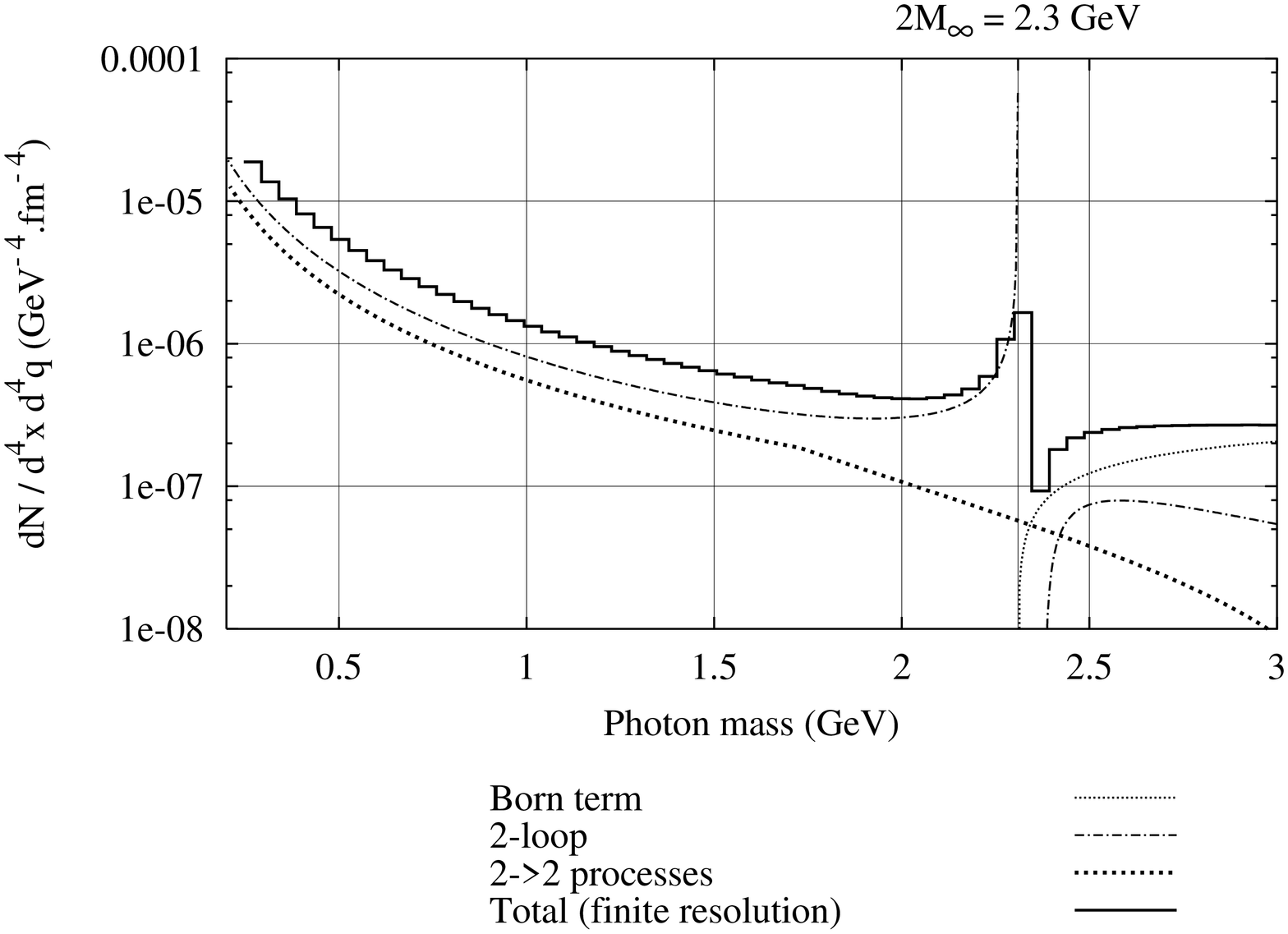}}}}
\caption{\label{fig:rate-mass} The dilepton rate in physical units, 
  for $N_{_{F}}=2$, $T=1$ GeV, $g=2$ and a fixed energy $q_0=5$ GeV, as a
  function of the mass of the pair.}
\end{figure}

\begin{figure}
\centerline{\resizebox*{!}{7cm}{\rotatebox{-90}{\includegraphics{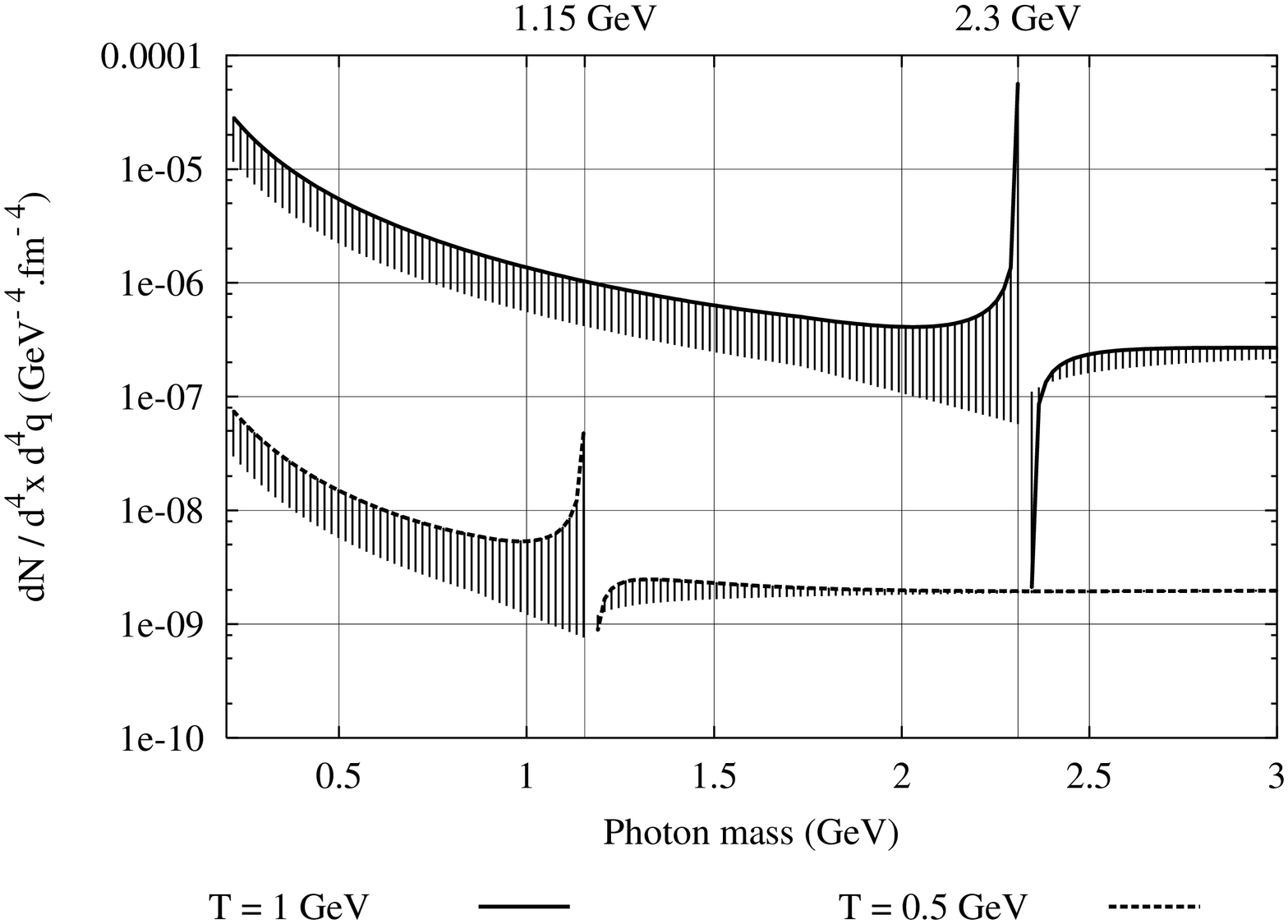}}}}
\caption{\label{fig:rate-T} The dilepton rate in physical units, 
  for $N_{_{F}}=2$, $g=2$ and a fixed energy $q_0=5$ GeV, as a
  function of the mass of the pair, for two values of the temperature
  $T=1$~GeV and $T=0.5$~GeV. The vertical lines indicate the
  contribution of the terms calculated in the present paper to the
  rate.}
\end{figure}

One can plot first the mass spectrum, at a fixed photon energy. In figure
\ref{fig:rate-mass}, the photon energy is set to $q_0=5$ GeV, and we
plot the yields for masses between $200$ MeV and $3$ GeV.
It appears that at this value of the energy, the 2-loop processes we
compute in this paper is comparable or even slightly larger than the
1-loop HTL contribution, especially in the vicinity of the threshold.
At the same energy of $5$ GeV, this process is slightly more important
for a lower temperature, as illustrated in figure \ref{fig:rate-T}.

The other important remark is that at such values of the coupling
constant, the thermal masses are not small, and the threshold for the
Born process is located at rather high masses. Therefore, most of the
spectrum is in fact dominated by formally higher order terms.

\subsection{Energy spectrum}
\begin{figure}
\centerline{\resizebox*{!}{7cm}{\rotatebox{-90}{\includegraphics{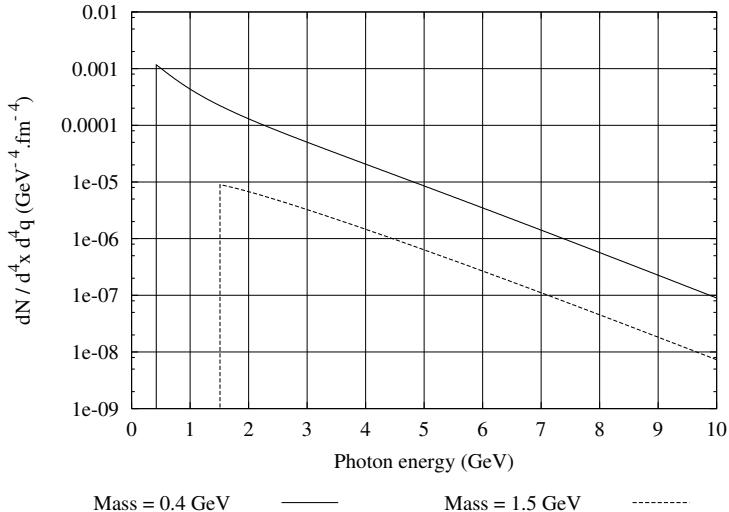}}}}
\caption{\label{fig:rate-energy} The dilepton rate in physical units, 
  for $N_{_{F}}=2$, $T=1$ GeV, $g=2$ and at two fixed pair masses
($Q=400$ MeV and $Q=1.5$ GeV), as a function of the energy of the
pair. All the contributions of figure \ref{fig:rate-mass} are added up.}
\end{figure}

In figure \ref{fig:rate-energy}, we set the photon
mass to some fixed value ($Q=400$ MeV and $Q=1.5$ GeV), and we plot
the dilepton yield as a function of the energy, for energies between
$q_0=\sqrt{Q^2}$ and $10$ GeV.
 Note that in the region where $q_0\sim Q$, the approximations
made in this paper as well as in \cite{AltheR1} for the $2\to 2$
processes are a priori not valid. We observe a very fast drop of the
yield with energy, following the usual exponential law in
$\exp(-q_0/T)$.

\section{Conclusions}
In this paper, we have calculated the contribution of bremsstrahlung and
off-shell annihilation to the production of a high energy low mass
dilepton by a quark gluon plasma. As long as the photon mass remains
small in front of its energy, this process is collinearly enhanced in
the same way as for the production of a real photon. Cancellations
between real and virtual cuts ensure that all infrared or collinear
divergences cancel. 

Interestingly, our result display a general feature of perturbative
expansions: such an expansion usually breaks down in the immediate
vicinity of tree-level phase-space boundaries. Practically, this means
that more work (read resummations) is needed near the tree-level
threshold ($Q^2=4M_\infty^2$) in order to calculate accurately these
processes.

Concerning phenomenology, we find again that the off-shell annihilation
is dominant for very large dilepton energies. This new contribution
therefore enhances the thermal dilepton rates at moderate invariant
masses and large energies, and it would be interesting to include it in
hydrodynamical evolution codes in order to see whether it leads to
visible effects in a realistic heavy-ion collision scenario.

\section*{Acknowledgments}
P.A. and F.G. would like to thank V. Ruuskanen and G.D. Moore for
useful discussions. H.Z. thanks LAPTH for hospitality during the
summer of 2001, where part of this work was done. F.G. thanks the
ECT*, where part of this work has been performed, for hospitality and
support.

\appendix

\section{Integration over $p_\perp^2$}
\label{app:integral}
\subsection{Case $M_{\rm eff}^2>0$}
This is the simplest case since the complex conjugation in
Eq.~(\ref{eq:ImPi}) merely brings a factor 2. Indeed, the quantities
$u+M_{\rm eff}^2$ and $(u+v+M_{\rm eff}^2)^2-4uv$ never vanish for
positive $u$ and $v$. In addition, we always have $u+v+M_{\rm
  eff}^2>0$. In this case, the only nontrivial integral we need to
perform is
\begin{equation}
A\equiv\int\limits_0^{+\infty}\!
{{du}\over{u+M_{\rm eff}^2}}{{1}\over{\left[(u+v+M_{\rm
eff}^2)^2-4uv\right]^{1/2}}}+{\rm c.c.}\; .
\end{equation}
This integral is elementary. First rewrite $(u+v+M_{\rm
eff}^2)^2-4uv=(u-v+M_{\rm eff}^2)^2+4vM_{\rm eff}^2$. Then, introduce a
new variable $t$ defined by
\begin{equation}
\ln(t)\equiv {\rm sinh}^{-1}\left(
{{u-v+M_{\rm eff}^2}\over{2M_{\rm eff}\sqrt{v}}}
\right)\; ,
\end{equation}
which brings the integral $A$ to the form
\begin{equation}
A={2\over{M_{\rm eff}\sqrt{v}}}\int\limits_{M_{\rm
eff}/\sqrt{v}}^{+\infty}{{dt}\over{t^2+{{\sqrt{v}}\over{M_{\rm eff}}}t-1}}\; ,
\end{equation}
at which point the integration is trivial:
\begin{equation}
A={2\over{\sqrt{v(v+4M_{\rm eff}^2)}}}\ln\left(
{{v+2M_{\rm eff}^2+\sqrt{v(v+4M_{\rm eff}^2)}}\over
{v+2M_{\rm eff}^2-\sqrt{v(v+4M_{\rm eff}^2)}}}
\right)\; .
\end{equation}
At this stage, it is sufficient to use the relation
\begin{equation}
2{\rm tanh}^{-1}(x)={\rm tanh}^{-1}\left(
{{2x}\over{1+x^2}}
\right)
\end{equation}
in order to obtain the first of Eqs.~(\ref{eq:int-u-basic}).

\subsection{Case $M_{\rm eff}^2<0$}
This case is more complicated that the previous one, mainly because
$\Delta\equiv(u+v+M_{\rm eff}^2)^2-4uv$ and $u+v+M_{\rm eff}^2$ can both
become negative. Let us first note that $1/\sqrt{\Delta}$ should in fact
be interpreted as\footnote{The sign in front of the imaginary part is
controlled by the sign of the infinitesimal imaginary part of $\overline{M_{\rm
eff}^2}$.}
\begin{equation}
{1\over{\sqrt{\Delta}}}={{\theta(\Delta)}\over{\sqrt{\Delta}}}-i\,{\rm sign}(p_0r_0){{\theta(-\Delta)}\over{\sqrt{-\Delta}}}\; .
\end{equation}
Similarly, we must write:
\begin{equation}
{1\over{u+\overline{M_{\rm eff}^2}}}={{\cal P}\over{u+M_{\rm eff}^2}}-i\pi\,{\rm
sign}(p_0r_0)\delta(u+M_{\rm eff}^2)\; .
\end{equation}
Therefore, we have for the integral $A$ the following expression
\begin{equation}
A=2\int\limits_0^{+\infty}\!\!du\left[{{\theta(\Delta)\,{\rm
sign}(u+v+M_{\rm eff}^2)}\over{\sqrt{\Delta}}}{{\cal P}\over{u+M_{\rm eff}^2}}
-\pi\delta(u+M_{\rm eff}^2)
{{\theta(-\Delta)}\over{\sqrt{-\Delta}}}
\right]\; .
\label{eq:def-A}
\end{equation}
One notes that in this equation, the second term corresponds to the
cuts denoted $(a)$ and $(c)$ in figure \ref{fig:diagrams}. This is the
reason why this term is always zero if $Q^2<4M_\infty^2$.

\subsubsection{$v+4M_{\rm eff}^2>0$}
Let us note first that this integral simplifies if
$v>-4M_{\rm eff}^2$ since then $\Delta$ and $u+v+M_{\rm eff}^2$ are both
positive at all $u$. Using now the following change of variable
\begin{equation}
\ln(t)\equiv {\rm cosh}^{-1}\left(
{{u-v+M_{\rm eff}^2}\over{2M_{\rm eff}\sqrt{v}}}
\right)\; ,
\label{eq:change-var}
\end{equation}
and employing a strategy similar to the one used
before, we find easily
\begin{equation}
A_{v+4M_{\rm eff}^2>0}={8\over{\sqrt{v(v+4M_{\rm eff}^2)}}}
{\rm tanh}^{-1}\sqrt{{v+4M_{\rm eff}^2}\over v}\; ,
\end{equation}
which proves the first of Eqs.~(\ref{eq:A-neg}).

\subsubsection{$v+4M_{\rm eff}^2<0$}
Things are more involved if $v+4M_{\rm eff}^2<0$ since now both $\Delta$
and $u+v+M_{\rm eff}^2$ can become negative. The second term in
Eq.~(\ref{eq:def-A}) is now different from zero and is given by:
\begin{equation}
A_2=-2\pi{{\theta(-v-4M_{\rm eff}^2)}\over{\sqrt{-v(v+4M_{\rm
eff}^2)}}}\; .
\label{eq:cut-cut}
\end{equation}
In order to calculate the first term of Eq.~(\ref{eq:def-A}), we need to
study the zeros of $\Delta$ as well as the sign of $u+v+M_{\rm
eff}^2$. We find that $\Delta$ is positive for $u\in[0,v-M_{\rm
eff}^2-2\sqrt{-vM_{\rm eff}^2}]$ or $u\in[v-M_{\rm
eff}^2+2\sqrt{-vM_{\rm eff}^2},+\infty[$. For $u>v-M_{\rm
eff}^2+2\sqrt{-vM_{\rm eff}^2}$, one can check that $u+v+M_{\rm eff}^2$
is always positive. On the contrary, for $u\in[0,v-M_{\rm
eff}^2-2\sqrt{-vM_{\rm eff}^2}]$, $u+v+M_{\rm eff}^2$ is positive if
$v>-M_{\rm eff}^2$ and negative if $v<-M_{\rm eff}^2$.  For
$u\in[v-M_{\rm eff}^2+2\sqrt{-vM_{\rm eff}^2},+\infty[$, the appropriate
change of variable is again the one given in Eq.~(\ref{eq:change-var}),
which leads to the following contribution to $A$:
\begin{eqnarray}
&& 2\int\limits_{v-M_{\rm eff}^2+2\sqrt{-vM_{\rm
eff}^2}}^{+\infty}\!\!\!{{du}\over{u+M_{\rm
eff}^2}}{{1}\over{\left[(u+v+M_{\rm
eff}^2)^2-4uv\right]^{1/2}}}\nonumber\\  &=&
{2\over{\sqrt{-v(v+4M_{\rm eff}^2)}}}\left[{\pi\over
2}-{\rm tan}^{-1}\sqrt{{v}\over{-v-4M_{\rm eff}^2}}\,\right]\; .
\label{eq:A1}
\end{eqnarray}
For $u\in[0,v-M_{\rm eff}^2-2\sqrt{-vM_{\rm
eff}^2}]$, one must define instead
\begin{equation}
\ln(t)\equiv {\rm cosh}^{-1}\left(-
{{u-v+M_{\rm eff}^2}\over{2M_{\rm eff}\sqrt{v}}}
\right)\; .
\end{equation}
By the same method, one can check that
\begin{eqnarray}
 &&2\int\limits_0^{v-M_{\rm eff}^2-2\sqrt{-vM_{\rm
eff}^2}}\!\!\!{{du}\over{u+M_{\rm eff}^2}}{{{\rm sign}(u+v+M_{\rm eff}^2)}\over{\left[(u+v+M_{\rm
eff}^2)^2-4uv\right]^{1/2}}}\nonumber\\
&=&{2\over{\sqrt{-v(v+4M_{\rm eff}^2)}}}\left[{\pi\over
2}-3{\rm tan}^{-1}\sqrt{{v}\over{-v-4M_{\rm eff}^2}}\,\right]\; .
\label{eq:A2}
\end{eqnarray}

Adding up the three contributions of Eqs.~(\ref{eq:cut-cut}),
(\ref{eq:A1}) and (\ref{eq:A2}), we obtain the following result for $A$
in the region where $v+4M_{\rm eff}^2<0$:
\begin{equation}
A_{v+4M_{\rm eff}^2<0}=-{8\over{}\sqrt{-v(v+4M_{\rm eff}^2)}}
{\rm tan}^{-1}\sqrt{{v\over{-v-4M_{\rm eff}^2}}}\; ,
\end{equation}
which proves the second of Eqs.~(\ref{eq:A-neg}). It is instructive to
note that we have a cancellation between the first and second term of
Eq.~(\ref{eq:def-A}) of a term that would have been singular in the
limit of zero momentum transfer ($v\to 0$). This is nothing but a
manifestation of the KLN theorem for the thermal production of a massive
particle, since these two terms correspond respectively to real and
virtual cuts. Therefore, this is yet another example to fuel the
controversy between \cite{KapusW1,KapusW2,KapusW3} and \cite{AurenBBGG1},
which does not support the finding by
\cite{KapusW1,KapusW2,KapusW3} that the Kinoshita-Lee-Nauenberg
\cite{Kinos1,LeeN1} theorem is invalid. Note that the present
verification of the KLN theorem, for bremsstrahlung and off-shell
annihilation processes, involves diagrams that in fact appear for the
first time at three loops in the bare perturbative expansion.

\bibliographystyle{unsrt} 

\end{document}